\documentclass[aps,prd,superscriptaddress,twocolumn, preprintnumbers,nofootinbib,10pt]{revtex4-2}
\usepackage{multirow}
\usepackage{amsmath}
\usepackage{amssymb}
\usepackage[dvipdf,dvips]{graphicx}
\usepackage{color}
\usepackage{hyperref}
\usepackage{url}
\usepackage{slashed}
\usepackage{subfigure}
\usepackage[usenames,dvipsnames]{xcolor}
\usepackage{amsmath}
\usepackage{amsfonts}
\usepackage{float} 
\usepackage{amssymb}
\usepackage{epsfig}
\usepackage{graphics}
\usepackage{euscript}
\usepackage{slashed}
\usepackage{epstopdf}
\usepackage[utf8]{inputenc}
\allowdisplaybreaks
\usepackage[normalem]{ulem}
\usepackage{pifont}
\usepackage{dsfont}
\usepackage{MnSymbol}
\usepackage{graphicx}
\usepackage{latexsym}
\usepackage{tikz-feynman}
\usepackage{tikz-cd}
\usepackage{easyReview}
\usepackage{cancel}
\usepackage[normalem]{ulem}
\usepackage{svg}

\makeatletter
\DeclareRobustCommand{\cev}[1]{%
	{\mathpalette\do@cev{#1}}%
}
\newcommand{\do@cev}[2]{%
	\vbox{\offinterlineskip
		\sbox\z@{$\m@th#1 x$}%
		\ialign{##\cr
			\hidewidth\reflectbox{$\m@th#1\vec{}\mkern4mu$}\hidewidth\cr
			\noalign{\kern-\ht\z@}
			$\m@th#1#2$\cr
		}%
	}%
}
\makeatother

\begin{document}

\title{Non-Abelian Casimir energy in the Curci-Ferrari model through a functional approach}


\author{David Dudal}
\email{david.dudal@kuleuven.be} 
\affiliation{KU Leuven Campus Kortrijk–Kulak, Department of Physics,	Etienne Sabbelaan 53 bus 7657, 8500 Kortrijk, Belgium}
	
\author{Philipe De Fabritiis} 
\email{pdf321@cbpf.br}
\affiliation{CBPF $-$ Centro Brasileiro de Pesquisas Físicas, Rua Doutor Xavier Sigaud 150, 22290-180, Rio de Janeiro, Brazil}

\author{Sebbe Stouten}
\email{sebbe.stouten@kuleuven.be} 
\affiliation{KU Leuven Campus Kortrijk–Kulak, Department of Physics,	Etienne Sabbelaan 53 bus 7657, 8500 Kortrijk, Belgium}

	
\begin{abstract}
		
Using functional integral methods, we investigate the non-Abelian Casimir energy in the Curci-Ferrari model, which offers an effective description of the infrared regime of Yang-Mills theory. We consider a 3+1D (resp.\ 2+1D) system of two infinite parallel plates (resp.\ wires) at a fixed distance from each other, with either perfect magnetic conductor (PMC) or perfect electric conductor (PEC) boundary conditions. Imposing the boundary conditions directly in the functional integral by the introduction of suitable auxiliary fields that act as Lagrange multipliers, we obtain a boundary effective action that captures the dynamics of this system. The Casimir energy is then computed both directly from the functional integral and via the energy-momentum tensor, providing equivalent results. We find that the Casimir energy for PEC and PMC conditions differs by a constant factor, which can be traced back to a van Dam--Veltman--Zakharov-like discontinuity  (both in 3+1D and 2+1D). Lastly, we show that our analytical results are compatible with a variety of recent numerical lattice simulations of the non-perturbative Yang-Mills Casimir energy, in which a novel non-perturbative mass scale emerges.
		
\end{abstract}

\maketitle	

\section{Introduction}	\label{SecIntro}

The Casimir effect~\cite{Casimir:1948dh} stands as a remarkable manifestation of the quantum vacuum fluctuations, allowing us to bridge the gap between the theoretical foundations of Quantum Field Theory (QFT) and a measurable phenomenon observed in laboratory experiments. Its most emblematic example is the attractive force between two electrically neutral, perfectly conducting, parallel flat plates. Although the attraction between neutral but polarizable bodies had been already explained by London in 1930~\cite{London30}, Casimir's approach was different. In fact, he computed the force between the plates through the difference in the zero-point energy of the electromagnetic field due to the presence of the non-trivial boundaries~\cite{Casimir:1948dh}. A conclusive experimental confirmation of the Casimir effect came only in 1997~\cite{Lamoreaux1997} but nowadays there have been done many other precise experiments~\cite{Bimonte:2021sib}. For some excellent reviews about the Casimir effect, see Refs.~\cite{Plunien:1986ca, Bordag:2001qi, Milton:2004ya, Bordag:2009zz}

Recently, a novel approach to study the Casimir effect using boundary QFT techniques was proposed in Ref.~\cite{Dudal:2020yah}, inspired by the works~\cite{Bordag:1985rb,Golestanian_1998}. The idea is to use functional integral methods and impose the boundary conditions through auxiliary fields that act as Lagrange multipliers. Integrating out the gauge fields, we can construct an effective boundary action for the auxiliary fields living on the boundary that codifies the information about the gauge fields and can be used to compute the vacuum energy directly through a functional determinant. This offers a flexible and elegant framework, that allows us to investigate the Casimir effect in scenarios that can go far beyond the usually considered ones. For recent developments using this approach, see Refs.~\cite{Canfora:2022xcx, Oosthuyse:2023mbs, Dudal:2024PEMC, Dudal:2024Robin, Dudal:2024DEM}.

The non-Abelian version of the Casimir effect opens a fascinating window into the interplay between the quantum vacuum structure and the confinement problem in Quantum Chromodynamics (QCD), one of the most challenging open questions nowadays~\cite{Greensite11}. Roughly speaking, the color-charged particles entering the strong interactions (quarks and gluons) cannot be directly observed. Considering pure Yang-Mills theories (YM), a linear confining potential between test charges was found in lattice simulations~\cite{Wilson:1974sk,Bali:2000gf,Lucini:2001nv,Luscher:2002qv}. Such a linear confining potential could be explained by the formation of a chromo-electric flux tube between the charges, in close resemblance with the formation of magnetic flux tubes in superconductors, defining the dual superconductor picture of the QCD vacuum~\cite{Mandelstam:1974pi,Nambu74,tHooft:1981bkw,Baker:1991bc}. From this perspective, studying the Casimir effect in non-Abelian gauge theories could offer clues about the underlying mechanism responsible for flux tube formation and the QCD vacuum.

A simple idea to phenomenologically model the QCD vacuum is to use the MIT bag model~\cite{Chodos74}, where the constituent quarks are locked in a color-neutral bag (implemented with suitable boundary conditions) that will correspond to the QCD bound state observables. The Casimir effect could provide a mechanism to guarantee the bag stability, but adopting standard perturbative techniques within the YM action one is led to a Casimir force with the ``wrong" sign~\cite{Boyer68,Davies72,Bender76}, destabilizing the bag instead of holding its constituents together. Nevertheless, it was shown in Refs.~\cite{Oxman05,Canfora13} that by adopting a non-perturbative propagator for the gluon, one can find an attractive Casimir force, showing that the infrared dynamics of YM theories play an important role.
 
One promising avenue to circumvent these challenges is to introduce a controlled modification of the infrared behavior by incorporating an effective mass scale that reflects the non-perturbative features of the theory. A particularly compelling implementation of this idea is provided by the Curci-Ferrari (CF) model~\cite{Pelaez21, Tissier10, Tissier11, CurciFerrari76}, the simplest renormalizable extension of the Faddeev-Popov Lagrangian that has the same field content, preserves standard perturbation theory in the UV, and describes a massive behavior for the gluon propagator in the infrared agreeing with lattice simulation results. The CF model stands as a simple effective description of the infrared regime of QCD, allowing us to perform reliable yet manageable analytical calculations. For recent developments using the CF model, see Refs.~\cite{Oribe:2025ezp, Barrios:2024ixj, Pelaez:2022rwx, Barrios:2022hzr, Barrios:2021cks}. An excellent pedagogical review can be found in Ref.~\cite{Reinosa:2024vph}.

Recent results for the non-Abelian Casimir effect in lattice simulations~\cite{Chernodub:2023dok,Ngwenya:2025cuw,Ngwenya:2025mpo} reveal a nontrivial behavior of the Casimir energy in the presence of spatial boundaries, characterized by the emergence of a new and unexpected intrinsic mass scale. This emergent mass scale suggests novel non-perturbative features, raising fundamental questions about its origin, analytical description, and its possible relation with confinement. Furthermore, there are other first-principles numerical simulations showing that boundaries can affect non-perturbative properties, as one can see in Refs.~\cite{Chernodub:2018pmt, Chernodub:2019nct, Chernodub:2017mhi, Chernodub:2022izt}.

The main goal of this work is to investigate the non-Abelian Casimir effect using the novel functional integral techniques developed in Ref.~\cite{Dudal:2020yah}, applied to the CF model~\cite{Pelaez21} as a suitable effective description of the infrared dynamics of YM theory, providing a fresh look through analytical computations in a boundary QFT approach. We compute the Casimir energy using functional methods and the energy-momentum tensor, aiming to better understand the non-perturbative vacuum structure of YM theories.

Perhaps surprisingly, we find that the Casimir energy for perfect electric conductor (PEC) and perfect magnetic conductor (PMC) boundary conditions differs by a factor \(3/2\). This difference originates from a van Dam--Veltman--Zakharov-like discontinuity, manifested through the presence (or absence) of a residual gauge symmetry at the level of the effective action of the auxiliary fields. Because of these residual gauge symmetries, PMC plates in the massive theory allow for an extra degree of freedom compared to both PMC plates in the massless theory and PEC plates in both the massive and massless cases, thus explaining the factor. 
Lastly, we explore how these analytical results compare with recent lattice data, seeking to comprehend the reported intrinsic mass scale~\cite{Chernodub:2023dok}.

This paper is organized as follows. We present our theoretical framework in Sec.~\ref{SecTheorySetup}, describing the model, the geometry, and the boundary conditions adopted here. In Sec.~\ref{SecBQFT}, we discuss the boundary QFT approach and use it to compute the Casimir energy in 3+1D through functional methods. In Sec.~\ref{SecEMT}, we follow another route and compute the Casimir energy using the energy-momentum tensor, showing that both methods give the same result. This serves as a good internal consistency check. Having done these computations in detail for PMC plates, in Sec.~\ref{SecPEC} we briefly describe the results for PEC plates. In Sec.~\ref{SecDiscont} we discuss the difference in Casimir energy for PMC and PEC plates, together with its discontinuity in the massless limit. The functional method described above can readily be applied to the case of parallel wires in 2+1D, which is discussed in Sec.~\ref{SecWire}. We compare our findings with recent lattice data in Sec.~\ref{SecLattice} and state our concluding remarks in Sec.~\ref{SecConclusions}. 

\section{Theoretical setup}\label{SecTheorySetup}

Let us consider the CF model~\cite{Pelaez21} with a general gauge parameter $\xi$ in a four-dimensional Euclidean spacetime, whose Lagrangian is given by:
\begin{align}
    \mathcal{L}_{\rm CF} &= \frac{1}{4} F_{\mu \nu}^a F_{\mu \nu}^a + \frac{1}{2} \partial_\mu \bar{c}^a \left(D_\mu c\right)^a + \frac{1}{2} \left(D_\mu \bar{c}\right)^a  \partial_\mu c^a \nonumber \\
    &+\frac{\xi}{2} h^a h^a + i h^a \partial_\mu A_\mu^a - \xi \frac{g^2}{8} \left(f^{abc} \bar{c}^b c^c\right)^2 \nonumber \\
    &+\frac{m^2}{2} A_\mu^a A_\mu^a + \xi m^2 \bar{c}^a c^a,
\end{align}
where $F_{\mu \nu}^a = \partial_\mu A_\nu^a - \partial_\nu A_\mu^a + g f^{abc} A_\mu^b A_\nu^c$ is the non-Abelian field strength, $D_\mu^{ab} = \delta^{ab} \partial_\mu - g f^{abc} A_\mu^c$ is the covariant derivative in the adjoint representation and $f^{abc}$ are the structure constants of $SU(N)$, with $a= 1, ..., N^2-1$. The fields $\left(c, \bar{c}\right)$ are the Faddeev-Popov ghosts, $h$ is the Nakanishi-Lautrup field used to enforce the gauge-fixing condition, and $m$ is a phenomenologically-motivated mass parameter introduced after the gauge-fixing.

We are interested in the leading contribution to the Casimir energy, thus we can restrict ourselves to terms quadratic in the fields. Since we will ignore interaction terms, we can safely neglect ghost contributions since they will not have any information about the boundaries in the absence of interactions. Integrating out the Nakanishi-Lautrup field, we obtain
\begin{align}
    S_{\rm CF}' = \!\!\int \!\! d^4x \left[\frac{1}{4} \mathcal{F}_{\mu \nu}^a \mathcal{F}_{\mu \nu}^a + \frac{1}{2 \xi} \left(\partial_\mu A_\mu^a\right)^2 + \frac{m^2}{2} A_\mu^a A_\mu^a \right],
\end{align}
where we defined the quadratic approximation for the non-Abelian field strength tensor as $\mathcal{F}_{\mu \nu}^a \equiv \partial_\mu A_\nu^a - \partial_\nu A_\mu^a$. The Landau gauge limit $\xi\to0$ is implicitly understood.

The geometry adopted here is a pair of parallel infinitely large and thin plates at a distance $L$ from each other, located at $z = z_\pm \equiv \pm L/2$. We define $\Sigma_{\pm}$ as the 3-dimensional subspaces defined by the plates at $z = z_\pm$ and the unit vector normal to $\Sigma_\pm$ by $n_\mu = \left(0,0,0,1\right)$.

Inspired by the dual superconductor picture of the QCD vacuum~\cite{Mandelstam:1974pi,Nambu74,tHooft:1981bkw,Baker:1991bc}, let us consider the PMC boundary conditions\footnote{In Sec.~\ref{SecPEC} we will consider PEC boundary conditions, but first we will perform all the computations in the PMC case to describe the method in detail without distractions.}, defined by 
\begin{align}\label{PMC}
	\mathcal{F}_{\mu \nu}^a n_\nu  \bigg\vert_{\Sigma_\pm} = 0.
\end{align}
This is a gauge invariant condition. The same applies to the PEC condition of Eq.~\eqref{pecvgl} to be discussed later.

Defining $G_\mu^a \equiv  \mathcal{F}_{\mu \nu}^a n_\nu$, we can rewrite the above boundary condition as $G_\mu^a\big\vert_{\Sigma_\pm} = 0$. One can impose these boundary conditions directly in the functional integral through the inclusion of the following term in the action:
\begin{align}\label{SBC-1}
        S_{\rm BC} = \int_{\Sigma^-} d^3 \bold{x} \, b_i^{-,a} (\bold{x}) G_i^a + \int_{\Sigma^+} d^3 \bold{x} \, b_i^{+,a} (\bold{x}) G_i^a,
    \end{align}
where $\left(b_i^{-,a}(\bold{x}), b_i^{+,a}(\bold{x})\right)$ are boundary auxiliary fields that depend only on $\bold{x} = (t,x,y)$, and only have Latin indices $(i = 0,1,2)$ since $G_3^a \equiv 0$ automatically. One can retrieve the boundary conditions defined in Eq.~\eqref{PMC} from the action~\eqref{SBC-1} via the equations of motion of the boundary auxiliary fields. 

For later convenience, we define the following antisymmetric tensor (here the index $\gamma \in \{ -, + \}$):
\begin{align}
H_{\mu \nu}^a = \delta(z-z^\gamma) \, \left(b_\mu^{\gamma, a} n_\nu - b_\nu^{\gamma, a} n_\mu\right).    
\end{align}
This allows us to write the boundary condition enforcing term in a manifestly covariant form
\begin{align}\label{SBC}
S_{\rm BC} = \int \! d^4x \, \frac{1}{2} H_{\mu \nu}^a \mathcal{F}_{\mu \nu}^a.
\end{align}
Including this boundary condition enforcing term, the quadratic action that will be used in this work is
\begin{align} \label{LagCFbound2}
    S = S_{\rm CF}' + S_{\rm BC}.
\end{align}
We remark that, although the mass term spoils the standard BRST symmetry, the CF model enjoys a modified BRST symmetry which is not nilpotent, but recovers the standard one in the $m \rightarrow 0$ limit~\cite{Pelaez21}. The modified BRST transformation that leaves the CF action invariant is given by (see Ref.~\cite[Sec.~4.2]{Pelaez21}) 
\begin{align}
    s A_\mu^a &= \left(D_\mu c\right)^a, \nonumber \\
    s c^a &= -\frac{g}{2} f^{abc} c^b c^c, \nonumber \\
    s \bar{c}^a &= i h^a, \nonumber \\
    s(ih^a) &= m^2 c^a.
\end{align}
The inclusion of \(S_\text{BC}\) does not modify the BRST properties of the theory. Indeed, if one defines the BRST transformation of the auxiliary fields to be \(sb_i^{\gamma,a} =0\), then one has that \(sS_\text{BC} = 0\) (since \(s \mathcal{F}_{\mu\nu}^a =0\) trivially).

The above action can be rewritten in Fourier space as\footnote{Technically, the non-trivial boundaries we are considering are interfaces/defects placed inside the bulk of unbounded Euclidean space, and not actual boundaries of the manifold. Therefore, in the action integral, one can integrate by parts and safely neglect total derivatives that vanish at infinity.}
\begin{align}
    S =  \! \int \!\!\! \frac{d^4 k}{(2 \pi)^4} \left[ \frac12 A_\mu^a(k) K^{ab}_{\mu \nu} A_\nu^b(-k) +v_\nu^a(k) A_\nu^a(-k) \right],
\end{align}
where we defined $ v_\nu^a(k) \equiv i k_\mu H_{\mu \nu}^a(k)$ and
\begin{align}\label{quad-operator-CF}
    K^{ab}_{\mu \nu} = \delta^{ab}\left[\left(k^2+m^2\right)\delta_{\mu \nu} - \left(1 - \frac{1}{\xi}\right) k_\mu k_\nu\right].
\end{align}
The operator $ K^{ab}_{\mu \nu}$ has a well-defined inverse, given by
\begin{align}\label{AdynOp2}
    \left(K^{-1}\right)^{ab}_{\mu \nu} &= \delta^{ab}\left[\frac{1}{k^2+m^2} \Theta_{\mu \nu} + \frac{\xi}{k^2 + \xi m^2} \Omega_{\mu \nu}\right],
\end{align}
where we defined the longitudinal and transverse projectors as $\Omega_{\mu \nu}(k) \equiv \frac{k_\mu k_\nu}{k^2}$ and $\Theta_{\mu \nu}(k) \equiv \delta_{\mu \nu} - \Omega_{\mu\nu}(k)$ respectively. Therefore, the functional integral that describes the system  can be written with the above action as:
\begin{align}
    Z = \int \! \mathcal{D}A_\mu \mathcal{D}b^+ \mathcal{D}b^- e^{-S}. 
\end{align}
In the following section, we will integrate out the gluon field to obtain an effective boundary theory, and compute the Casimir energy from the remaining functional integral.


\section{The boundary effective action and the Casimir energy}\label{SecBQFT}

First of all, notice that it is possible to decouple the boundary field sector from the gluon one by performing a suitable shift in the gauge field (that will not change the functional measure), whose form is given by: 
\begin{align}
A_\mu^a(k) \rightarrow A_\mu^a(k) - v_\rho^b(k) \left(K^{-1}\right)^{ab}_{\mu \nu}.
\end{align}
Indeed, after this shift, we find:
\begin{align} \label{Stotalaftershift}
    S = \frac{1}{2} \int \! \frac{d^4 k}{(2 \pi)^4} &\bigg\lbrace  A_\mu^a(k) K^{ab}_{\mu \nu} A_\nu^b(-k) \nonumber \\
    &- v_\mu^a(k) \left(K^{-1}\right)^{ab}_{\mu \nu} v_\nu^b(-k) \bigg\rbrace.
\end{align}
One can immediately see that the functional integral decouples. That is, one can write $Z = Z_A \, Z_b$, where the first part only contains the gluon field and the second part is written in terms of the auxiliary fields. Now, the gluon part does not have any dependence on the plate separation, and thus will not contribute to the Casimir effect. Indeed, after the shift, all the information about the geometric boundaries is contained in \(Z_b\), thus we will focus on it from now on, discarding \(Z_A\). 

There is a simple way to understand why we can simply discard contributions that do not depend on $L$ when computing the Casimir energy. The reason is that the Casimir energy of the system needs to be defined as the energy difference between the system with boundaries and the system without~\cite{Casimir:1948dh}. Note that, in practice, the latter energy can be obtained from the former one by simply taking the limit $L \rightarrow \infty$. Therefore, when considering the energy difference, any term independent of $L$ will vanish, being irrelevant for the Casimir energy.

The functional integral in the boundary field sector is
\begin{align}
    Z_b = \int \! \mathcal{D}b^+ \mathcal{D}b^- \, e^{-S_b},
\end{align}
where the boundary effective action $S_b$ can be obtained by using $v_\nu^a(k) \equiv i k_\mu H_{\mu \nu}^a(k)$ in Eq.~\eqref{Stotalaftershift}, giving us
\begin{align}
    S_b = -  \frac{1}{2} \! 
    \int \!\! \frac{d^4 k}{(2 \pi)^4} &\bigg\lbrace  k_\mu H_{\mu \rho}^a(k) \left(K^{-1}\right)^{ab}_{\rho \sigma} k_\nu H_{\nu \sigma}^b(-k) \bigg\rbrace.
\end{align}
We can write $H_{\mu \nu}^a$ in Fourier space as
\begin{align}
    H_{\mu \nu}^a(k) &= e^{i k_z z^\gamma} H_{i \mu \nu} b_i^{\gamma, a}(\bold{k}),
\end{align}
where
\begin{align}
    H_{i \mu \nu} \equiv \delta_{i \mu} n_\nu - \delta_{i \nu} n_\mu.
\end{align}
Thus, the boundary effective action can be rewritten as
\begin{align} \label{BEA2}
    S_b = -\frac{1}{2} \int &\frac{d^4k}{(2\pi)^4} \left[ e^{i k_z (z^\gamma - z^\lambda)}  b_i^{\gamma, a}(\bold{k}) b_j^{\lambda, b}(\bold{-k}) \right. \nonumber \\ 
    &\left. \times k_\mu k_\nu H_{i \mu \rho} H_{j \nu \sigma}\left(K^{-1}\right)_{\rho \sigma}^{ab}  \right].
\end{align}

We can simplify the term in the second line using the antisymmetric nature of the tensor $H_{i \mu \nu}$. Indeed, any contribution proportional to $k_\rho k_\sigma$ will vanish due to symmetry arguments. Thus, in this computation,  we can effectively set $\left(K^{-1}\right)_{\rho \sigma}^{ab} \longrightarrow  \frac{\delta^{ab} \delta_{\rho \sigma}}{k^2+m^2}$, giving us
\begin{align}
    &k_\mu k_\nu H_{i \mu \rho} H_{j \nu \sigma} \left(K^{-1}\right)_{\rho \sigma}^{ab}  =     \frac{\delta^{ab}\left[k^2 \delta_{ij} - \vert\bold{k}\vert^2 \Theta_{ij}(\bold{k})\right]}{k^2+m^2},
\end{align}
where $\Theta_{ij}(\bold{k})$ is the three-dimensional version of the transverse projector defined before. We remark that the gauge parameter $\xi$ naturally drops out at this point.

Thus, we obtain for the boundary effective action:
\begin{align}\label{Sbpartial2}
    &S_b = -\frac{1}{2} \int \!  \frac{d^3\bold{k}}{(2\pi)^3} \bigg\lbrace b_i^{\gamma,a}(\bold{k})  \, b_j^{\lambda,b}(-\bold{k}) \, \delta^{ab}\nonumber \\
    &\int \!\! \frac{dk_z}{2\pi} \left[e^{i k_z(z^\gamma- z^\lambda)}  \frac{k^2}{k^2 + m^2} \left( \delta_{ij} - \frac{\vert\bold{k}\vert^2}{k^2} \Theta_{ij}(\bold{k})\right)  \right] 
 \bigg\rbrace.
\end{align}

The next step is to perform the integral in $dk_z$. To accomplish this task, we will use the following integral formulas (valid for any $\omega_k > 0$ and $\Lambda \in \mathbb{R}_0$):
\begin{align} \label{IntegralFormulas}
    \int \! \frac{dk_z}{2\pi} \frac{e^{i k_z \Lambda}}{k_z^2 + \omega_k^2} &= \frac{e^{-\omega_k \vert \Lambda \vert}}{2 \omega_k}, \nonumber \\
     \int \! \frac{dk_z}{2\pi} \frac{k_z^2 \, e^{i k_z \Lambda}}{k_z^2 + \omega_k^2} &= -\frac{1}{2} \omega_k \, e^{-\omega_k \vert \Lambda \vert},
\end{align} 
where we defined $\omega_k \equiv \sqrt{\vert\bold{k}\vert^2 + m^2}$. For future use, we also collect the following useful integrals here:
\begin{align}  \label{IntegralFormulas2}
    \int \! \frac{dk_z}{2\pi} \frac{e^{i k_z \Lambda}}{\left(k_z^2 + \omega_k^2\right)^2} &= \frac{\left(1 + \omega_k \vert \Lambda \vert \right) e^{-\omega_k \vert \Lambda \vert}}{4 \omega_k^3}, \nonumber \\
     \int \! \frac{dk_z}{2\pi} \frac{k_z^2 \, e^{i k_z \Lambda}}{\left(k_z^2 + \omega_k^2\right)^2} &= \frac{\left(1 - \omega_k \vert \Lambda \vert \right) e^{-\omega_k \vert \Lambda \vert}}{4 \omega_k}, \nonumber \\
     \int \! \frac{dk_z}{2\pi} \frac{k_z^4 \, e^{i k_z \Lambda}}{\left(k_z^2 + \omega_k^2\right)^2} &= -\frac{1}{4} \omega_k\left(3 - \omega_k \vert \Lambda \vert \right) e^{-\omega_k \vert \Lambda \vert}.
\end{align}
In the above integrals, terms $\propto \delta(\Lambda)$ were always dropped as $\Lambda\neq0$. One can then check  that, either by direct integration or by using dimensional regularization in case of divergent integrals, that the (regularized) results also apply for the case $\Lambda=0$.

Thus, after performing the integral in $dk_z$, we find the following expression for the boundary effective action:
\begin{align} \label{SbcompactwithK}
    S_b = \frac{1}{2} \int \! \frac{d^3 \bold{k}}{(2\pi)^3} \, b_i^{\gamma, a} \,\mathbb{K}_{ij}^{\gamma \lambda}(\bold{k})  \, b_j^{\lambda, a},
\end{align}
where the dynamical operator $\mathbb{K}$ is given by
\begin{align}\label{KoperatorCF}
\mathbb{K}_{ij}^{\gamma \lambda}(\bold{k}) =\frac{\omega_k}{2} \left[ \Theta_{ij} + \left(\frac{m^2}{\omega_k^2}\right) \Omega_{ij} \right] e^{-\omega_k \vert z^\gamma - z^\lambda \vert}.
\end{align}
Therefore, we obtain a boundary effective action quadratic in the boundary fields, with a well-defined dynamical operator. Since this object is diagonal in color space, we can do all the computations ignoring the color structure and then multiply the final result by $N^2-1$. It is important to remark that in the limit $m \rightarrow 0$, we have $\omega_k \rightarrow \vert \bold{k} \vert$, thus recovering the known YM expression.

Before we start computing the Casimir energy using the boundary effective action, for future use, let us briefly compute the inverse of the boundary dynamical operator $\mathbb{K}$. Its $6\times6$ matrix structure can be written as 
\begin{align} \label{Kmatrixstructure}
\mathbb{K} = 
\begin{pmatrix}
    \mathbb{K}_{ij}^{++} & \mathbb{K}_{ij}^{+-} \\
    \mathbb{K}_{ij}^{-+} & \mathbb{K}_{ij}^{--} 
\end{pmatrix},
\end{align}
where each entry represents a $3\times3$ block, given by:
\begin{align} \label{Kmatrixblocks}
     \mathbb{K}_{ij}^{++} &= \mathbb{K}_{ij}^{--} = \frac{\omega_k}{2} \left[ \Theta_{ij} + \left(\frac{m^2}{\omega_k^2}\right) \Omega_{ij} \right], \nonumber \\
     \mathbb{K}_{ij}^{+-} &= \mathbb{K}_{ij}^{-+} = \frac{\omega_k}{2} \left[ \Theta_{ij} + \left(\frac{m^2}{\omega_k^2}\right) \Omega_{ij} \right] e^{- L \omega_k}.
\end{align}
The inverse of $\mathbb{K}$ can be written in a convenient form as
\begin{align} \label{KinvEandF}
    \left(\mathbb{K}^{-1}\right)_{ij}^{\gamma \lambda}(\bold{k}) = E^{\gamma\lambda}\Theta_{ij}(\bold{k}) +F^{\gamma\lambda} \Omega_{ij}(\bold{k}),
\end{align}
where \(\Theta_{ij}(\bold{k})\) and \(\Omega_{ij}(\bold{k})\) are the 3D transverse resp.~longitudinal projectors, and we have
\begin{align} \label{EandFofK}
    E^{\gamma\lambda} &= \frac{2}{\omega_k \left(1-e^{-2 L \omega_k}\right)} \begin{pmatrix}
        1 & -e^{-L \omega_k} \\
        -e^{-L \omega_k} & 1
    \end{pmatrix}, \\
   F^{\gamma\lambda} &= \frac{2 \omega_k}{m^2\left(1-e^{-2 L \omega_k}\right)} \begin{pmatrix}
        1 & -e^{-L \omega_k} \\
        -e^{-L \omega_k} & 1
    \end{pmatrix}.
\end{align} 
Now we can compute the Casimir energy using Eq.~\eqref{SbcompactwithK}.

It is well-known that from the functional integral, one can obtain the vacuum energy through the expression
\begin{align}
    Z = e^{-V \mathcal{E}},
\end{align}
where $V$ is the spacetime volume and $\mathcal{E}$ is the vacuum energy density. On the other hand, the functional integral associated with the boundary effective action is quadratic in the boundary fields, thus we can perform the Gaussian integral in these fields to obtain
\begin{align}
    Z_b = \int \! \mathcal{D}b^+ \mathcal{D}b^- \, e^{-S_b} = C \left({\rm Det} \mathbb{K}\right)^{-1/2},
\end{align}
where $C$ is an irrelevant infinite constant and $\mathbb{K}$ is given in Eq.~\eqref{KoperatorCF}. The functional determinant must be understood in a complete sense, including both the discrete and continuous indices. Using the well-known result $\log {\rm Det} A = {\rm Tr} \log A$, we find
\begin{align} \label{EnergyLogDet}
    \mathcal{E} = \frac{1}{2} \int \frac{d^3 \bold{k}}{(2\pi)^3} \log \left(\det \mathbb{K}(\bold{k})\right),
\end{align}
where $\det \mathbb{K}(\bold{k})$ denotes only the matrix determinant. 
Note that the quadratic operator \(\mathbb{K}\) in Eq.~\eqref{KoperatorCF} is a pure tensor product, meaning we can compute the determinant using the general identity
\begin{equation}
    \det(A_{n\times n}\otimes B_{m\times m}) = \left( \det A \right)^m \left( \det B\right)^n.
\end{equation}
This yields \(\det \mathbb{K}(\mathbf{k}) = \left(\frac{\omega_k}{2}\right)^6 \det \left[ \Theta_{ij} + \left(\frac{m^2}{\omega_k^2}\right) \Omega_{ij} \right]^2 (1-e^{-2L \omega_k})^3\). Since the eigenvector of \(\Omega\) is orthogonal to \(\Theta\), and the \((d-1)\) eigenvectors of \(\Theta\) are orthogonal to \(\Omega\), and all of them have eigenvalue \(1\), we get that \(\det (a \Theta + b \Omega) = a^{d-1}b.\) Applying this for \(d=3\) gives us the sought-after determinant
\begin{align}\label{DetKCF}
    \det \mathbb{K}(\mathbf{k}) = \frac{\omega_k^2 m^4}{2^6}  \, \left(1 - e^{-2 L \omega_k} \right)^3.
\end{align}

Alternatively, to compute the determinant of the block matrix $\mathbb{K}(\bold{k})$ (since $\mathbb{K}_{ij}^{--}$ is invertible), we can use the following expression:
 \begin{align} \label{detblock}
    \det 
     \begin{pmatrix}
     A & B \\
     C & D 
 \end{pmatrix} = \det \left[A - B D^{-1} C\right] \cdot \det [D].
 \end{align}
 All blocks have a factor $\frac{\omega_k}{2}$, thus this will give us a $\left(\frac{\omega_k}{2}\right)^6$ in the end. To compute the determinant of each $3 \times 3$ block, we can use the general formula for $3 \times 3$ matrices
 \begin{align} \label{dettrace}
     \det M= \frac{1}{6} \left({\rm tr} M\right)^3 - \frac{1}{2} {\rm tr} \left(M^2\right)  {\rm tr} M + \frac{1}{3} {\rm tr} \left(M^3\right).
 \end{align}
 Since all the blocks are written in terms of projectors, this computation can be easily performed, giving us once more~\eqref{DetKCF}.

We remark that at this point we cannot consider the $m \rightarrow 0$ limit straightforwardly because in the absence of a mass, the operator $\mathbb{K}_{ij}^{--}$ would not be invertible, jeopardizing the way we computed the determinant here. We will encounter related interesting behavior in the massless limit later on.

To compute the Casimir energy, terms that do not depend on $L$ are not relevant. Thus, inserting the above expression in the vacuum energy equation and discarding terms not depending on $L$, we find
\begin{align} \label{EnergyCFlogdet}
    \mathcal{E}_\text{PMC} &= \frac{3 (N^2-1)}{2} \int \frac{d^3 \bold{k}}{(2\pi)^3}   \log \left[1 - e^{-2 L \omega_k}\right] \nonumber \\
    &= \frac{3 (N^2-1)}{4 \pi^2} \int_0^\infty \! dk \,  k^2  \log \left[1 - e^{-2 L \sqrt{k^2 + m^2}}\right].
\end{align}
We will discuss this result in detail in Sec.~\ref{SecDiscont}.

\section{Energy-Momentum tensor method for the Casimir energy}\label{SecEMT}

There is another well-known method for computing the vacuum energy: through the energy-momentum tensor (EMT). In fact, since the EMT is a local object, this method is more flexible than the global method above, and thus more widely applicable to several extensions, e.g., to the dynamical Casimir effect for moving boundaries \cite{Bordag:1985rb}. Adopting this method, one notices that the EMT can have non-trivial contributions coming from the boundary if we consider more general boundary conditions \cite{Dudal:2024PEMC}, something that is rarely addressed in the literature. The starting point is given by Eq.~\eqref{LagCFbound2}, repeated here for convenience:
\begin{align} \label{StartLagEMT}
    S = \int \! d^4x &\left[ \frac{1}{4} \mathcal{F}_{\mu \nu}^a \mathcal{F}_{\mu \nu}^a + \frac{1}{2 \xi} \left(\partial_\mu A_\mu^a\right)^2 \right.\nonumber \\
    &\left.+ \frac{m^2}{2} A_\mu^a A_\mu^a + \frac{1}{2} H_{\mu \nu}^a \mathcal{F}_{\mu \nu}^a \right].
\end{align}
The equation of motion (EOM) in this approximation reads
\begin{align}\label{EOM}
    \partial_\rho \left(\mathcal{F}_{\mu \rho}^a  + H_{\mu\rho}^a \right) =   \frac{1}{\xi} \partial_\mu \partial_\alpha A_\alpha^a - m^2 A_\mu^a.
\end{align}
From the above equation, taking the derivative $\partial_\mu$ and using the antisymmetry of $\mathcal{F}_{\mu \rho}^a$ and $H_{\mu\rho}^a$, we can obtain $\partial_\alpha A_\alpha^a = 0$ as a subsidiary condition. Thus, when working on-shell, we can discard any contribution coming with a factor $\left(\partial A\right)$. Of course, this observation is a bit obsolete in our case since we are anyhow working in the Landau gauge, leading to the very same conclusion.

Here we will use the canonical EMT obtained through Noether's theorem~\cite{Noether1918}. For convenience, we sum a term $X_{\mu \nu} \equiv -\partial_\rho \left[\left(\mathcal{F}_{\mu \rho}^a +H_{\mu\rho}^a\right) A_\nu^a\right]$ that does not affect the EMT conservation since $\partial_\mu X_{\mu \nu} = 0$, thus being harmless. Therefore, we find for the EMT:
\begin{align}
    T_{\mu \nu}^{\rm CF} =  T_{\mu \nu}^{\rm YM} +  T_{\mu \nu}^{\rm mass} + T_{\mu \nu}^{\rm bnd},
\end{align}
where we organized the different contributions to the EMT according to their origin:
\begin{align} \label{EMTcontributions}
    T_{\mu \nu}^{\rm YM} &= \mathcal{F}_{\mu \rho}^a \mathcal{F}_{\nu \rho}^a - \frac{1}{4} \delta_{\mu \nu}   \mathcal{F}_{\alpha \beta}^a \mathcal{F}_{\alpha \beta}^a, \nonumber \\
     T_{\mu \nu}^{\rm mass} &= m^2 \left(A_\mu^a A_\nu^a - \frac{1}{2}\delta_{\mu \nu}  A_\alpha^a A_\alpha^a\right), \nonumber \\
    T_{\mu \nu}^{\rm bnd} &= H_{\mu \rho}^a \mathcal{F}_{\nu \rho}^a - \frac{1}{2}  \delta_{\mu \nu} H_{\alpha \beta}^a \mathcal{F}_{\alpha \beta}^a.     
\end{align}
From the EOM, remembering Eq.~\eqref{AdynOp2}, we can readily obtain
\begin{align}
    A_\sigma^a(p) = \left(K^{-1}\right)^{ab}_{\sigma \mu} \left(i p_\rho H_{\mu \rho}^b(p) \right) .
\end{align}
Taking into account that $H_{\mu \rho}^b(p)$ is antisymmetric in $\mu \rho$ and is already contracted with $p_\rho$, we can conclude that its contraction with any term proportional to $p_\mu$ will vanish. Thus, the EOM becomes
\begin{align} \label{EOMgluon}
    A_\sigma^a(p) =  \frac{i p_\rho}{p^2+m^2} H_{\sigma \rho}^a,
\end{align}
from which one can immediately find
\begin{align}
    \mathcal{F}_{\mu \nu}^a(p) = \left(\frac{1}{p^2 +m^2}\right) \left[p_\rho p_\mu H_{\nu \rho}^a - p_\rho p_\nu H_{\mu \rho}^a \right].  
\end{align}
The next step is to use the EOM to substitute the gluon fields appearing in the EMT to obtain an expression only containing the boundary fields.

For the record, if one feels uncomfortable to directly compute the EMT in Euclidean convention, the result can be checked upon starting from the Minkowski version of the CF action and rederive the EMT along similar lines, followed by a Wick rotation. This also explains the minus sign in the equation below.

Using the same reasoning as in Ref.~\cite[Sec.~IV.B]{Dudal:2024PEMC}, we write:
\begin{equation}
    \mathcal{E} = -\int\!dz\; \langle T^{\rm CF}_{00}\rangle = -\frac13 \int\!dz\; \langle T^{\rm CF}_{ii}\rangle.
\end{equation}
This allows us to avoid uncontracted indices, making the upcoming computations easier to perform. Thus, we can write the total Casimir energy as a sum of each contribution appearing in Eq.~\eqref{EMTcontributions}:
\begin{align}
    \mathcal{E} = \mathcal{E}^{\rm YM} + \mathcal{E}^{\rm mass} + \mathcal{E}^{\rm bnd}. 
\end{align}
Throughout this computation, some contractions will often appear, thus we write them here for convenience:
\begin{align}\label{eq:kkHH}
    k_\sigma k_\beta H_{i\rho\sigma} H_{j\rho\beta} &= k_z^2 \Theta_{ij} +k^2 \Omega_{ij}, \\
    k_\sigma k_\beta H_{ik\sigma} H_{jk\beta} &= k_z^2 \Theta_{ij} +k_z^2 \Omega_{ij},\\
    k_k k_\beta H_{i\rho k} H_{j\rho\beta} &=  \vert\bold{k}\vert^2 \Omega_{ij}.
\end{align}
Moreover, since we are expressing the EMT in terms of the boundary fields, in order to compute the vacuum expected value, we will need 
\begin{align}\label{VEV}
    \langle b_i^{\gamma, a}(\bold{k}) b_j^{\lambda, b}(\bold{q}) \rangle = (2 \pi)^3 \delta\left(\bold{k} + \bold{q}\right) \delta^{ab}\left(\mathbb{K}^{-1}\right)_{ij}^{\gamma \lambda}(\bold{k}).
\end{align}
Now we collect the different contributions to the EMT. 


From the YM term, we find two contributions, one coming from $\mathcal{F}_{i \rho}^a \mathcal{F}_{i \rho}^a$ and another from $ - \frac{1}{4} \delta_{ii}  \mathcal{F}_{\alpha \beta}^a \mathcal{F}_{\alpha \beta}^a$:
\begin{align} \label{EMTYM1}
\mathcal{E}^{\rm YM} =  \left(N^2-1\right) \!\! \int\!\!\frac{d^4k}{(2\pi)^4} \,  \frac{e^{i k_z (z^\gamma-z^\lambda)}}{\left(k^2 + m^2\right)^2}  \left[ X_1^{\gamma\lambda} + X_2^{\gamma\lambda} \right],
\end{align}
where 
\begin{align}
    X_1^{\gamma\lambda} &= \! -\frac{1}{3} \left[ \vert\bold{k}\vert^2 (2k_z^2 E^{\gamma\lambda} + k^2 F^{\gamma\lambda})+ k^2 k_z^2(2 E^{\gamma\lambda} + F^{\gamma\lambda}) \right], \nonumber \\
    X_2^{\gamma\lambda} &= \frac{k^2}{2} \left[ 2k_z^2 E^{\gamma\lambda} + k^2 F^{\gamma\lambda} \right].
\end{align}
Performing the integral in $dk_z$ using Eq.~\eqref{IntegralFormulas2} and plugging the explicit values for $E^{\gamma\lambda}$ and $F^{\gamma\lambda}$~\eqref{EandFofK}, we find:
\begin{align}
   \mathcal{E}^{\rm YM} &= - (N^2-1) \int\!\!\frac{d^3\bold{k}}{(2\pi)^3} \frac{1}{6 \omega_k^2} \times \nonumber \\
   &\left[ 3 \left( 4 \vert \bold{k} \vert^2 + 3 m^2 \right) + \frac{ L \omega_k \, e^{-2 L \omega_k} \left( 4 \vert \bold{k} \vert^2 + 3 m^2 \right) }{ \left(1 - e^{-2 L \omega_k}\right)} \right].
\end{align}
The first term does not have any dependence on $L$, thus it will not contribute to the Casimir energy and can be dropped. The relevant contribution to the Casimir energy is the second term: 
\begin{align}
   \mathcal{E}^{\rm YM} &= - (N^2-1) \!\! \int_{0}^{\infty}\!\!\!dk \, \frac{k^2 L \omega_k \, e^{-2 L \omega_k} \left( 4 k^2 + 3 m^2 \right)}{12 \pi^2 \omega_k^2 \left(1 - e^{-2 L \omega_k}\right)},
\end{align}
and one can see that it vanishes in the limit $L \rightarrow \infty$.
As another consistency check, one can consider the limit $m \rightarrow 0$ of this expression, to compare with the known expression for the vacuum energy for parallel plates, obtaining:
\begin{align}
   \mathcal{E}^{\rm YM}_{m \rightarrow 0} &= - \frac{(N^2-1)}{3 \pi^2} \!\! \int_{0}^{\infty}\!\!\!dk \, \left[\frac{k^3 L  \, e^{-2 L k} }{1 - e^{-2 L k}}\right] \nonumber \\
   &= (N^2-1) \left(-\frac{\pi^2}{720 L^3}\right).
\end{align}
That is, we obtain $(N^2-1)$ times the well-known result for the Maxwell Casimir energy with parallel plates, as expected.

Following the same reasoning, from the mass term we have contributions from $m^2 A_i^a A_i^a$ and $- \frac{m^2}{2}\delta_{ii}  A_\alpha^a A_\alpha^a$:
\begin{align} \label{EMTm1}
    \mathcal{E}^{\rm mass} \!=  &\left(N^2-1\right) \! \int \!\! \frac{d^4k}{(2\pi)^4} \,  \frac{e^{i k_z (z^\gamma-z^\lambda)}}{\left(k^2 + m^2\right)^2 } \left[ Y_1^{\gamma\lambda} + Y_2^{\gamma\lambda} \right],
\end{align}
where 
\begin{align}
    Y_1^{\gamma\lambda} &=  -\frac{m^2}{3}  k_z^2  \left(2 E^{\gamma\lambda} + F^{\gamma\lambda}\right), \nonumber \\
    Y_2^{\gamma\lambda} &= \frac{m^2}{2} \left(2k_z^2 E^{\gamma\lambda} + k^2 F^{\gamma\lambda}\right).
\end{align}
Using Eq.~\eqref{IntegralFormulas2} to integrate in $dk_z$, and substituting $E^{\gamma\lambda}$ and $F^{\gamma\lambda}$ from Eq.~\eqref{EandFofK}, we find
\begin{align}
    \mathcal{E}^{\rm mass} &= (N^2-1) \int\!\!\frac{d^3\bold{k}}{(2\pi)^3} \frac{1}{6 \omega_k^2} \left[ \left(4 \vert \bold{k}\vert^2 + 3 m^2\right)  \right. \nonumber \\
     &\left. + \frac{L \omega_k \, e^{-2L \omega_k}}{\left(1 - e^{-2L \omega_k}\right)}\left(3 m^2 - 2 \vert \bold{k}\vert^2\right)   \right].
\end{align}
Once again, the first term is independent of $L$ and the second one goes to zero in the limit $L \rightarrow \infty$. Again, discarding the \(L\)-independent first term, we obtain
\begin{align}
    \mathcal{E}^{\rm mass}  &= (N^2-1) \!\! \int_0^{\infty} \!\!\! dk \frac{k^2 L \omega_k  e^{-2L \omega_k} \left(3 m^2 - 2 k^2\right)}{12 \pi^2 \omega_k^2 \left(1 - e^{-2L \omega_k}\right)}.
\end{align}

Lastly, we have to find the boundary contribution. Here we also have two contributions, coming from $H_{i \rho}^a \mathcal{F}_{i \rho}^a$ and $ - \frac{1}{2}  \delta_{ii} H_{\alpha \beta}^a \mathcal{F}_{\alpha \beta}^a$, but this time the situation will be a bit different. Here we have:     
\begin{align} \label{EMTbnd1}
   \mathcal{E}^{\rm bnd}  =  \left(N^2-1\right) \int\!\frac{d^4k}{(2\pi)^4} \,  \frac{e^{i k_z (z^\gamma-z^\lambda)}}{k^2 + m^2}  \left[ Z_1^{\gamma \lambda} + Z_2^{\gamma \lambda} \right],
\end{align}
where
\begin{align}
    Z_1^{\gamma \lambda} &=  \frac{1}{3} \left[ \vert \bold{k}\vert^2  F^{\gamma\lambda} + k_z^2 (2 E^{\gamma\lambda} + F^{\gamma\lambda})\right], \nonumber \\
    Z_2^{\gamma \lambda} &= -\left( 2 k_z^2 E^{\gamma\lambda} + k^2 F^{\gamma\lambda}\right).
\end{align}
Performing the integral in $dk_z$ and substituting the explicit values of $ E^{\gamma \lambda}$ and $ F^{\gamma \lambda}$, we find
\begin{align}
      \mathcal{E}^{\rm bnd} = 4 (N^2-1) \int\!\!\frac{d^3\bold{k}}{(2\pi)^3} = 0.
\end{align}
This is an infinite constant that is zero in dimensional regularization (see e.g.~\cite[Eq.~(4.2.6)]{Collins:1984xc} and \cite[below Eq.~(10.9)]{Zinn-Justin:2002ecy}), and does not depend on $L$ anyway, so it can be dropped.
That is, for PMC boundary conditions, we find that the boundary terms do not contribute to the EMT for any value of $m$. However, it is important to remark that under more general boundary conditions, there can be a non-trivial boundary contribution to the EMT, as explicitly showed in Ref.~\cite{Dudal:2024PEMC}.

Having computed all contributions to the vacuum energy, we can sum them to find the final result for the Casimir energy for PMC plates:
\begin{align} \label{EnergyCFemtPMC}
    \mathcal{E}_\text{PMC} = -\frac{(N^2-1)}{2 \pi^2} \!\! \int_{0}^{\infty}\!\!\!dk \, \frac{k^4 L \, e^{-2 L \omega_k}}{\omega_k \left(1 - e^{-2 L \omega_k}\right)}.
\end{align}
At first glance, the final expression for the Casimir energy obtained through the EMT method, Eq.~\eqref{EnergyCFemtPMC}, seems to be different from the one obtained through the first method, Eq.~\eqref{EnergyCFlogdet}. However, after closer inspection, both expressions are related through an integration by parts. Therefore, we conclude that both methods are equivalent.

\section{Perfect electric conductor boundary conditions}\label{SecPEC}

So far, we have examined in detail the non-Abelian Casimir energy for parallel plates at a distance $L$ from each other, considering PMC boundary conditions. Now we discuss the PEC boundary conditions. The entire reasoning can be followed {\it mutatis mutandis}, thus, we will only report the main differences here. 

The PEC boundary conditions are defined by 
\begin{align}
    \tilde{\mathcal{F}}_{\mu \nu}^a n_\nu \bigg\vert_{\Sigma_\pm} = 0, \label{pecvgl}
\end{align}
where the dual field strength is $\tilde{\mathcal{F}}_{\mu \nu}^a = \frac{i}{2} \epsilon_{\mu \nu \rho \sigma} \mathcal{F}_{\rho \sigma}^a$ in Euclidean space\footnote{The factor $i$ shows up because we have Wick-rotated the Levi-Civita tensor to work in Euclidean space.}, with \(\epsilon\) the Levi-Civita symbol. As before, we can introduce a boundary term in the action~\eqref{SBC}, but this time the relevant antisymmetric tensor is 
\begin{align}
    H_{\mu \nu}^a = \delta(z - z^\gamma) i \epsilon_{i \mu \nu z} b_i^{\gamma, a}.
\end{align}
As in the PMC case, we can decouple the boundary field sector from the gluon sector, finding the same general structure for the boundary effective action, given by Eq.~\eqref{BEA2}, with a different antisymmetric tensor, now given by
\begin{align}
    H_{i \mu \nu} = i \epsilon_{i \mu \nu z}.
\end{align}
One can massage the boundary effective action using the same symmetry argument as before, allowing us to write:
\begin{align}
    &S_b = \frac{1}{2} \int \!  \frac{d^3\bold{k}}{(2\pi)^3} \bigg\lbrace b_i^{\gamma,a}(\bold{k})  \, b_j^{\lambda,b}(-\bold{k}) \, \delta^{ab}\nonumber \\
    &\int \!\! \frac{dk_z}{2\pi} \left[e^{i k_z(z^\gamma- z^\lambda)}  \frac{\vert\bold{k}\vert^2 \Theta_{ij}(\bold{k})}{k^2 + m^2}  \right] 
 \bigg\rbrace.
\end{align}
The $dk_z$ integral is simpler here. We can write the final expression for the effective boundary action in the form of Eq.~\eqref{SbcompactwithK}, but for PEC boundary conditions we have: 
\begin{align}
    \mathbb{K}_{ij}^{\gamma \lambda}(\bold{k}) = \frac{\vert\bold{k}\vert^2 \Theta_{ij}(\bold{k})}{2 \omega_k} e^{-\omega_k \vert z^\gamma - z^\lambda \vert}.
\end{align}
The transverse structure of the $ \mathbb{K}$ operator means there is a residual gauge symmetry $b_i^\gamma \rightarrow b_i^\gamma + \beta^\gamma k_i$. This can be interpreted as a gauge invariance of the boundary fields that must be fixed in order to provide meaningful results, and is a phenomenon already present in Maxwell theory \cite{Dudal:2020yah}. Adding the gauge-fixing terms $b_i^\pm \Omega_{ij} b_j^\pm$, we obtain:
\begin{align}\label{KoperatorPEC}
\mathbb{K}_{ij}^{\gamma \lambda}(\bold{k}) = \frac{\vert\bold{k}\vert^2}{2 \omega_k} \left[\Theta_{ij}(\bold{k}) + \eta \, \delta^{\gamma \lambda} \Omega_{ij}(\bold{k}) \right] e^{-\omega_k \vert z^\gamma - z^\lambda \vert},
\end{align}
where $\eta$ is a gauge parameter and there is no summation over the $\gamma,\lambda$ indices. Note that this corresponds to a common linear covariant gauge. Although not needed for the current purposes, this can be accommodated for in a full-fledged BRST invariant fashion from which the $\eta$-independence will follow.

The next step is to compute the Casimir energy from the identity $Z = e^{-V \mathcal{E}}$. Thus, we need to consider Eq.~\eqref{EnergyLogDet} with the PEC $\mathbb{K}$ operator given by Eq.~\eqref{KoperatorPEC}. Computing the determinant in a similar fashion as before, we get
\begin{align}\label{DetPEC}
    \det \mathbb{K} = \left(\frac{\vert \bold{k} \vert^2}{2 \omega_k}\right)^6 \eta^2 \left(1 - e^{-2 \omega_k L}\right)^2.
\end{align}
Discarding terms that do not depend on $L$ and multiplying by $(N^2-1)$, we find for PEC boundary conditions:
\begin{align} \label{EnergyCFlogdetPEC}
    \mathcal{E}_\text{PEC} = \frac{(N^2-1)}{2 \pi^2} \int_0^\infty \! dk \,  k^2  \log \left[1 - e^{-2 L \sqrt{k^2 + m^2}}\right].
\end{align}
For now, we note that the non-Abelian Casimir energy for parallel plates is different for PMC and PEC boundary conditions. Indeed, the energy expressions \eqref{EnergyCFlogdet} resp.~\eqref{EnergyCFlogdetPEC} have the same analytical structure, but differ by a multiplicative factor $\frac{3}{2}$. We will return to the interpretation of this factor in Sec.~\ref{SecDiscont}.

Let us now discuss the EMT method for PEC boundary conditions. The starting point is still given by Eq.~\eqref{StartLagEMT}. However, under PEC boundary conditions we have $H_{i \mu \nu} = i \epsilon_{i \mu \nu z}$, thus the contractions reported in Eq.~\eqref{eq:kkHH} will change accordingly. Notice that  $H_{i \mu \nu}$ vanishes if it has any index $z$ due to the Levi-Civita tensor, resulting in only a single contraction, given by $k_a k_b H_{i\rho a} H_{j\rho b} = - \vert\bold{k}\vert^2 \Theta_{ij}(\bold{k})$, which significantly simplifies the computation. Contracting it with the operator $ \mathbb{K}^{-1}$ written as in Eq.~\eqref{KinvEandF}, we find
\begin{align} \label{KcontractionPEC}
    k_\alpha k_\beta H_{i \rho \alpha} H_{j \rho \beta} (\mathbb{K}^{-1})_{ij}^{\gamma\lambda} = - 2 \vert\bold{k}\vert^2 E^{\gamma \lambda}.
\end{align}
That is, only the transverse part of $\mathbb{K}^{-1}$ will be relevant. Inverting the PEC dynamical operator~\eqref{KoperatorPEC}, we obtain
\begin{align}
    E^{\gamma \lambda} = \frac{2 \omega_k}{\vert \bold{k}\vert^2 \left(1- e^{-2  L\omega_k}\right)}  \begin{pmatrix}
        1 & -e^{-L \omega_k} \\
        -e^{-L \omega_k} & 1
    \end{pmatrix}. 
\end{align}

Now we compute the different EMT contributions. The expressions for $\mathcal{E}^{\rm YM}$, $\mathcal{E}^{\rm mass}$, and $\mathcal{E}^{\rm bnd}$ are still given by Eqs.~\eqref{EMTYM1},\eqref{EMTm1},\eqref{EMTbnd1}, but now with coefficients:
\begin{align}
    X_1^{\gamma\lambda} + X_2^{\gamma\lambda} &= \frac{2}{3} \vert\bold{k}\vert^2 E^{\gamma \lambda} \left( 2 \vert\bold{k}\vert^2 + k_z^2 \right)  - \vert\bold{k}\vert^2 E^{\gamma \lambda} \left(\vert\bold{k}\vert^2 + k_z^2\right), \nonumber \\
    Y_1^{\gamma\lambda} + Y_2^{\gamma\lambda} &= \frac{2 m^2}{3}  \vert \bold{k}\vert^2 E^{\gamma\lambda}  - m^2 \vert \bold{k}\vert^2 E^{\gamma\lambda}, \nonumber \\
     Z_1^{\gamma \lambda} + Z_2^{\gamma \lambda} &=  -\frac{2}{3} \vert \bold{k}\vert^2  E^{\gamma\lambda} + \vert \bold{k}\vert^2  E^{\gamma\lambda}.
\end{align}
The final result for each contribution is given by
\begin{align}
\mathcal{E}^{\rm YM} \! &= -  \!\!\!\int\!\!\! \frac{d^3\bold{k}}{(2\pi)^3} \frac{(N^2-1)}{ 3 \omega_k^2}\left[m^2 \!+\! \frac{e^{-2L \omega_k } \, L \omega_k \left(2 \vert \bold{k} \vert^2 + m^2\right)}{1 - e^{-2 L \omega_k}} \right], \nonumber \\
 \mathcal{E}^{\rm mass} &= -(N^2-1) \int\!\!\frac{d^3\bold{k}}{(2\pi)^3} \frac{m^2}{3 \omega_k^2} \left[ 1 - \frac{L \omega_k \, e^{-2 L \omega_k}}{1 - e^{-2L \omega_k}} \right], \nonumber \\
\mathcal{E}^{\rm bnd} &= \frac{2}{3} (N^2-1) \int\!\!\frac{d^3\bold{k}}{(2\pi)^3} = 0. 
\end{align}
Summing all these contributions and discarding the $L$-independent part, we obtain the Casimir energy: 
\begin{align} \label{EnergyCFemtPEC}
    \mathcal{E}_\text{PEC} = -\frac{(N^2-1) }{3 \pi^2}\!\! \int_{0}^{\infty}\!\!\!dk \, \frac{k^4 L  \, e^{-2 L \omega_k} }{\omega_k \left(1 - e^{-2 L \omega_k}\right)}.
\end{align}
Here again, performing an integration by parts, this expression can be brought into the exact same form as the result \eqref{EnergyCFlogdetPEC} from the other method. 

\begin{figure}[t!]
	\begin{minipage}[b]{1.0\linewidth}
\includegraphics[width=\textwidth]{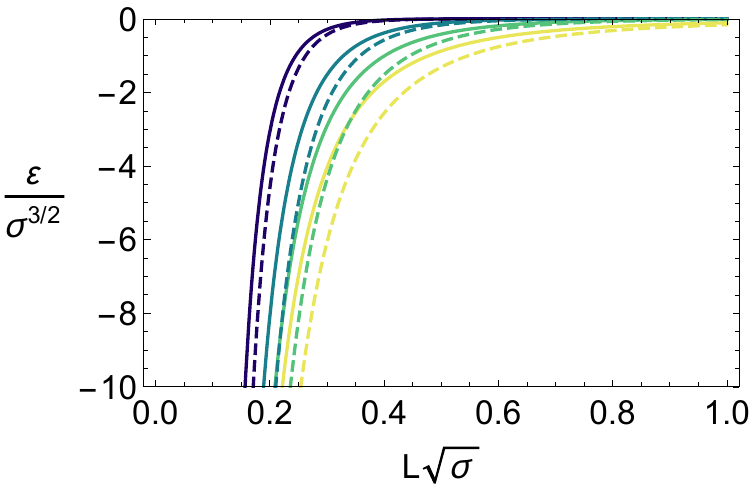}
	\end{minipage} \hfill
\caption{Analytic result for non-Abelian Casimir energy between parallel plates under PEC (full lines, \eqref{EnergyCFlogdetPEC}) and PMC (dashed lines, \eqref{EnergyCFlogdet}) boundary conditions for $SU(3)$ gauge group in four dimensions in units of $\sqrt{\sigma}$. Different values of the mass \(m\in \{0^+, 2, 4, 8\} \times \sqrt \sigma\) are represented in different colors, lighter colors corresponding to lower masses.}
\label{fig:Casimir4dSU3}
\end{figure}

A graphical representation of the analytic 
Casimir energy density for PEC~\eqref{EnergyCFlogdetPEC} and PMC~\eqref{EnergyCFlogdet} plates is shown in Fig.~\ref{fig:Casimir4dSU3}.

\section{Discontinuous Casimir energy in the massless limit}\label{SecDiscont}

As we have noted above, the non-Abelian Casimir energy for PMC and PEC plates differs by a numerical factor: \(\mathcal{E}_\text{PMC} = \frac32 \mathcal{E}_\text{PEC}\). This raises an interesting puzzle in the \(m\rightarrow0\) limit. Inspecting \(\mathcal{E}_\text{PEC}\) in Eq.~\eqref{EnergyCFlogdetPEC}, one immediately finds that its massless limit equals the Maxwell Casimir energy (accounted for color multiplicity): 
\begin{equation}
    \lim_{m\rightarrow0}\mathcal{E}_\text{PEC} = (N^2-1) \left(-\frac{\pi^2}{720 L^3}\right) \equiv \mathcal{E}_\text{Maxwell}.
\end{equation}
This means that we get a discontinuity of the PMC Casimir energy in the massless limit: \(\lim_{m\rightarrow0}\mathcal{E}_\text{PMC} = \frac32 \mathcal{E}_\text{Maxwell}\), whereas for \(m\) exactly zero (i.e.~Maxwell theory), we obviously have \(\mathcal{E}_\text{PMC}^{m=0} = \mathcal{E}_\text{Maxwell}\), without the factor \(3/2\).

This phenomenon of having a discontinuous massless limit in a gauge theory has already been observed in several settings and is known under the name \emph{van Dam--Veltman--Zakharov discontinuity}, or vDVZ discontinuity \cite{vanDam:1970vg,Zakharov:1970cc}. Intuitively, this effect can be traced back to the discrete difference in the number of degrees of freedom of a massless/massive spin 1 particle. However, in the case of the non-Abelian Casimir energy, this cannot be the full story, since the vDVZ-like discontinuity only occurs for PMC plates, and not for PEC plates. Moreover, we should also keep in mind that the CF model is defined including its (Landau) gauge fixing, it is not a massive gauge theory that is gauge fixed {\it a posteriori} (that is, we are not considering massive Yang-Mills). 

In order to better understand the different behavior of the two boundary conditions, let us look for the precise origin of the vDVZ-like discontinuity we are encountering here. Thanks to the two methods we have at our disposal for calculating the Casimir energy, we can look at it from two different angles. 

In the direct functional integral method, the difference between PMC and PEC plates is the presence of a residual gauge freedom. For PMC plates, the effective boundary theory \eqref{SbcompactwithK} does not exhibit a residual gauge freedom, i.e.~the quadratic form \(\mathbb{K}\) is invertible. For PEC plates on the other hand, the effective boundary theory does have one degree of residual gauge freedom, i.e.~the quadratic form \(\mathbb{K}(\mathbf{k})\) is only of rank 2. This residual gauge freedom effectively eats away one degree of freedom, reducing the power of \(\left(1 - e^{-2 L \omega_k} \right)^3\) in the PMC determinant \eqref{DetKCF} to \(\left(1 - e^{-2 L \omega_k} \right)^2\) in the PEC determinant \eqref{DetPEC}, leading directly to the factor of \(3/2\). These two degrees of freedom are also what we expect for a massless gauge boson in (3+1)D, hence the continuous massless limit for the PEC case. We will therefore speak about a vDVZ-like discontinuity.

In the EMT method, the origin of the vDVZ-like discontinuity can be traced back to the vacuum expectation value (VEV) of the mass contribution \(T_{\mu\nu}^\text{mass}\) to the EMT. Indeed, both for PMC and PEC we have that \(T_{\mu\nu}^\text{mass}\rightarrow0\) when \(m\rightarrow0\), but after taking the VEV this is not true anymore. For PEC, we do still have that \(\mathcal{E}^\text{mass}\rightarrow0\) after the VEV; but for PMC \(\mathcal{E}^{\rm mass}\rightarrow \frac{1}{2} \mathcal{E}_\text{Maxwell}\), accounting for the full vDVZ-like discontinuity. Since taking the VEV involves the inverse quadratic form \(\mathbb{K}^{-1}\) (cfr.~\eqref{VEV}), we again trace back the vDVZ-like discontinuity to residual gauge freedom of the effective boundary theory.

The natural question that arises is: can the presence of a vDVZ-like discontinuity already be discerned from the original 4D action, without having to inspect the effective boundary action? For this, consider a residual gauge transformation \(b\rightarrow b+\partial\phi\) at the level of the 4D action. Then for PMC, the boundary action term \eqref{SBC-1} transforms schematically as \(b n \mathcal{F} \rightarrow bn\mathcal{F} - \phi n \partial \mathcal{F}\) after partial integration. For PEC, we analogously have \(b n \tilde{\mathcal{F}} \rightarrow bn\tilde{\mathcal{F}} - \phi n \partial\tilde{\mathcal{F}}\). However, because of the Bianchi identity for \(\tilde{\mathcal{F}}\), we always have that \(\partial\tilde{\mathcal{F}} =0\). This means that for PEC plates, there is a residual gauge symmetry for any \(m\). For PMC though, the EOM \eqref{EOM} on the plates yields \(\partial \mathcal{F} = m^2 A\).\footnote{
There is a subtlety regarding the boundary term in the EOM. Since the boundary condition is only applied to the plates, the boundary action term is actually \(\delta(z-z^\gamma)b n \mathcal{F} \rightarrow \delta(z-z^\gamma) \left( bn\mathcal{F} - \phi n \partial \mathcal{F}\right) \). The EOM is given by \(\partial\mathcal{F} = m^2 A - \partial H\), with \(H\) also living on the plates, carrying a factor \(\delta(z-z^\lambda)\) itself. If the plate indices \(\gamma\neq \lambda\), then the product \(\delta(z-z^\gamma)\delta(z-z^\lambda)=0\), and if the plates indices \(\gamma=\lambda\), then the product \(\delta(z-z^\gamma)\delta(z-z^\lambda)\) yields a \(\delta(0)\) after integration, which is zero as well in dimensional regularization. So it is indeed allowed to simply use the EOM \(\partial\mathcal{F} = m^2 A\) on the plates.}
So for \(m\neq 0\), there is no residual gauge symmetry for PMC. This nicely explains the discrete difference between the massive PMC case on the one side, and the massless PMC case and both the massive and massless PEC case on the other side.


\section{The parallel wire case in (2+1)D} \label{SecWire}

\begin{figure}[t!]
	\begin{minipage}[b]{1.0\linewidth}
\includegraphics[width=\textwidth]{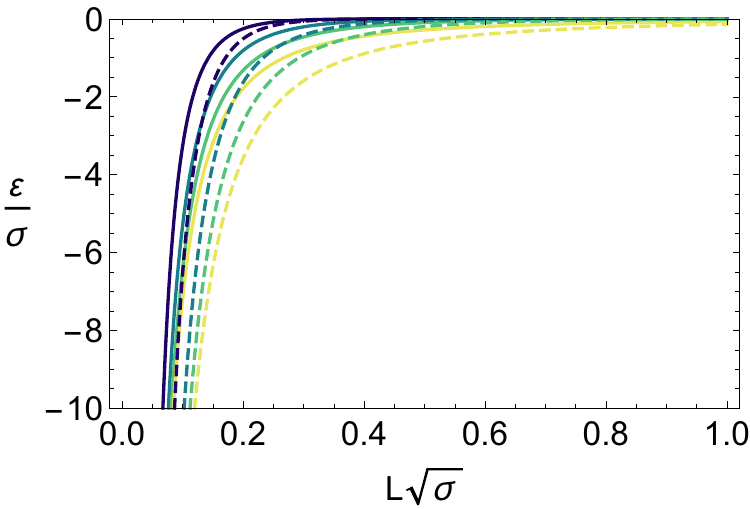}
	\end{minipage} \hfill
\caption{Analytic result for non-Abelian Casimir energy between parallel wires under PEC (full lines, \eqref{CasimirWirePEC}) and PMC (dashed lines, \eqref{CasimirWirePMC}) boundary conditions for $SU(2)$ gauge group in three dimensions in units of $\sqrt{\sigma}$. Different values of the mass \(m\in \{0^+, 2, 4, 8\} \times \sqrt \sigma\) are represented in different colors, lighter colors corresponding to lower masses.}
\label{fig:Casimir3dSU2}
\end{figure}

Let us consider now a simpler situation, where we have two infinitely long and thin parallel wires at a distance $L$ from each other in a three-dimensional Euclidean space. This is the (2+1)-dimensional counterpart of the parallel plates setup studied in this work. The theoretical setup adopted in this section will be the same as before, besides the obvious change in the geometry. Computing the Casimir energy in this scenario for PMC and PEC boundary conditions would follow exactly the same steps, that will not be repeated here. We will focus on the main changes, adopt only the functional method, state the results and compare with lattice data in the following.

Considering PMC boundary conditions, independently of the dimensionality, the relevant dynamical boundary operator is given by Eq.~\eqref{KoperatorCF}, which we repeat here:
\begin{align}
\left(\mathbb{K}^\text{PMC}\right)_{ij}^{\gamma \lambda}(\bold{k}) =\frac{\omega_k}{2} \left[ \Theta_{ij} + \left(\frac{m^2}{\omega_k^2}\right) \Omega_{ij} \right] e^{-\omega_k \vert z^\gamma - z^\lambda \vert}.  \nonumber
\end{align}
Its determinant, in the parallel wire case for PMC boundary conditions can be written as
\begin{align}
    \det \mathbb{K}^{\rm PMC} = \frac{m^4}{16} \left(1 - e^{-2 L \omega_k}\right)^2.
\end{align}
We can use this expression to compute the Casimir energy using Eq.~\eqref{EnergyLogDet}, discarding terms that do not depend on $L$, as usual, and including the color factor $(N^2-1)$:
\begin{align}
    \mathcal{E}_{\rm wire}^{\rm PMC} &=  \frac{(N^2-1)}{2\pi} \int_0^\infty \! dk \,  k \log \left(1 - e^{-2 L \omega_k}\right).
\end{align}
This time, the above integral has a closed expression:
\begin{align} \label{CasimirWirePMC}
 \mathcal{E}_{\rm wire}^{\rm PMC} \!\!=\! -\frac{(N^2-1)}{8 \pi L^2} \! \left[ 2 L m  \, {\rm Li}_2 \! \left(e^{-2 L m}   \right) \!+\! {\rm Li}_3 \!\left(e^{-2 L m} \right) \right],
\end{align}
where the polylogarithm function is: ${\rm Li}_n(z) = \sum_{k=1}^\infty \frac{z^k}{k^n}$. This functional form was also found in \cite{Karabali:2018ael}.

Now, we move on and consider PEC boundary conditions. The relevant operator for this case is Eq.~\eqref{KoperatorPEC}:
\begin{align}
\left(\mathbb{K}^\text{PEC}\right)_{ij}^{\gamma \lambda}(\bold{k})  = \frac{\vert\bold{k}\vert^2}{2 \omega_k} \left[\Theta_{ij}(\bold{k}) + \eta \, \delta^{\gamma \lambda} \Omega_{ij}(\bold{k}) \right] e^{-\omega_k \vert z^\gamma - z^\lambda \vert}. \nonumber
\end{align}
Computing its determinant, this time we get
\begin{align}
    \det \mathbb{K}^{\rm PEC} = \left(\frac{\vert \bold{k}\vert^2}{2 \omega_k}\right)^4 \eta^2 (1 - e^{-2 L \omega_k}).
\end{align}
Using Eq.~\eqref{EnergyLogDet} and discarding terms independent of $L$:
\begin{align}
    \mathcal{E}_{\rm wire}^{\rm PEC} 
    &=  \frac{(N^2-1)}{4 \pi} \int_0^\infty \! dk \,  k \log \left(1 - e^{-2 L \omega_k}\right).
\end{align}
Comparing the PEC and PMC expressions, we see that 
\begin{align} \label{CasimirWirePEC}
     \mathcal{E}_{\rm wire}^{\rm PEC} = \frac{1}{2}  \mathcal{E}_{\rm wire}^{\rm PMC}.
\end{align}
The massless limit can be readily considered for the wires:
\begin{align}
\lim_{m \rightarrow 0} \mathcal{E}_{\rm wire}^{\rm PEC} = - (N^2-1) \frac{\zeta(3)}{16 \pi L^2},
\end{align}
where $\zeta(3) \approx 1.20$ is the so-called Apéry's constant. The above result agrees with the one obtained in Ref.~\cite{Dudal:2020yah} for parallel wires in the Abelian massless case (ignoring the color factor, of course). Also here there is a mismatch in the Casimir energy for PMC and PEC boundary conditions (this time a factor of 2). Once again we can interpret this discrepancy as coming from a difference in the number of degrees of freedom for each boundary condition, due to an extra gauge symetry in the PEC case.

A graphical representation of the analytic Casimir energy density for PEC~\eqref{CasimirWirePEC} and PMC~\eqref{CasimirWirePMC} wires is shown in Fig.~\ref{fig:Casimir3dSU2}.

\begin{figure}[t!]
	\begin{minipage}[b]{1.0\linewidth}
\includegraphics[width=\textwidth]{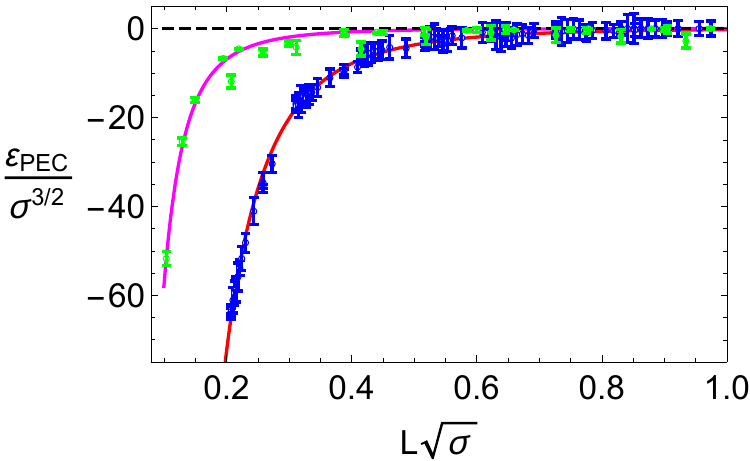}
	\end{minipage} \hfill
\caption{Non-Abelian Casimir energy between parallel plates under PEC boundary conditions for $SU(3)$ gauge group in four dimensions in units of $\sqrt{\sigma}$. The red and magenta curves represent the analytical expression~\eqref{CasimirEnergyCF}, using $m=0.54 \, {\rm GeV}$ as input and global factors $C = 5.63$ and $C = 0.54$ to fit the lattice data of~\cite{Chernodub:2023dok} (in blue) and~\cite{Ngwenya:2025cuw} (in green), respectively.}
\label{CasimirCF4dSU3Joint}
\end{figure}

\begin{figure}[t!]
	\begin{minipage}[b]{1.0\linewidth}
\includegraphics[width=\textwidth]{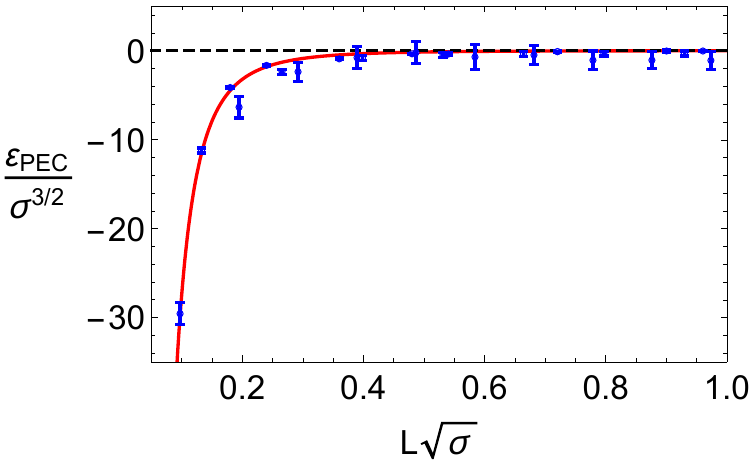}
	\end{minipage} \hfill
\caption{Non-Abelian Casimir energy between parallel plates under PEC boundary conditions for $SU(2)$ gauge group in four dimensions in units of $\sqrt{\sigma}$. The red curve represents the analytical expression~\eqref{CasimirEnergyCF}, using $m=0.68 \, {\rm GeV}$ as input and global factor $C = 0.69$ to fit the lattice data of~\cite{Ngwenya:2025cuw} (in blue).}
\label{CasimirCF4dSU2}
\end{figure}

\section{Comparing with lattice data} \label{SecLattice}

The non-Abelian Casimir energy has been computed using first-principles numerical simulations in lattice gauge theory for different systems over the last few years~\cite{Chernodub:2023dok, Chernodub:2018pmt, Ngwenya:2025cuw,Ngwenya:2025mpo, Chernodub:2019nct}. It is essential to understand whether the analytical results obtained through the CF model allow us to obtain good agreement with these recent developments of lattice gauge theory. In the following, we will analyze separately the cases of parallel plates and wires. We have collected all lattice data in single graphs, we refer to the original works for details about the simulations themselves, the considered volumes, lattice spacings, etc.

\subsection{Parallel plates}

The Casimir energy between perfect chromometallic parallel plates in zero-temperature $SU(3)$ lattice gauge theory in $(3+1)$ dimensions was investigated recently in Ref.~\cite{Chernodub:2023dok}. In this work, the authors note that since in $(2+1)$ dimensions the non-Abelian Casimir energy can accurately be described as the Casimir energy of a massive scalar particle~\cite{Karabali:2018ael}, this assumption might as well lead to a good fit in $(3+1)$ dimensions. Therefore, they use the following expression \cite{Cougo-Pinto:1994obn} to fit the lattice data:
\begin{align}\label{CasimirDataFit}
    \mathcal{E}_{\rm data} = -C_0 \frac{(N^2-1) m_{\rm gt}^2}{4 \pi^2 L} \sum_{n=1}^{\infty} \frac{1}{n^2} K_2\left(2 n m_{\rm gt} L\right),
\end{align}
where $K_2$ is the modified Bessel function of the second kind and the best-fit values are $C_0 = 5.60(7)$ and $m_{\rm gt} = 0.49(5) \, {\rm GeV}$. It is emphasized that this emergent gluon mass scale is smaller than the mass of the lightest glueball, given by $M_{0^{++}} = 1.653~{\rm GeV}$~\cite{Athenodorou:2020ani}. Furthermore, the authors advocate that the glueball mass $M_{0^{++}}$ determines the mass gap in the bulk while this new mass scale emerges due to the existence of a non-trivial boundary, calling this boundary state a ``glueton" and interpreting it as a non-perturbative colorless state of gluons~\cite{Chernodub:2023dok}. The global parameter $C_0$, as introduced in~\cite{Chernodub:2023dok}, is phenomenological in nature and amasses all missing effects.   

The Casimir energy between parallel plates with PEC boundary conditions in the CF model is given by Eq.~\eqref{EnergyCFlogdetPEC}. Expanding the logarithm as a series, after a few technical steps similar to \cite[App.~B]{Canfora:2022xcx}, one can find:\footnote{Naturally, if we were adopting PMC boundary conditions, we would obtain $3/2$ times the above result.}
\begin{align}\label{CasimirEnergyCF}
     \mathcal{E}_{\rm PEC} = -\frac{ (N^2-1) m^2}{4 \pi^2 L} \sum_{n=1}^{\infty} \frac{1}{n^2}  K_2(2 n m L).
\end{align}
The expression~\eqref{CasimirEnergyCF} we obtain from the CF model only differs by a numerical factor from the proposed fit~\eqref{CasimirDataFit}. Therefore, the analytical result obtained from the CF model can surely fit the lattice data~\cite{Chernodub:2023dok} if we set $m = m_{\rm gt}$ and adopt a global factor $C = C_0$. 

In the same vein, we can also consider the recent lattice data of Refs.~\cite{Ngwenya:2025cuw,Ngwenya:2025mpo}, where the authors investigated the non-Abelian Casimir energy for many different systems in $(3+1)$ and $(2+1)$ dimensions with $SU(3)$ and $SU(2)$ gauge groups at zero temperature. In Ref.~\cite{Ngwenya:2025cuw}, the authors fitted their lattice data with a different functional form, viz.~$\sim L^{-3-\nu}e^{-m_{\rm gt}^\prime L}$, where $\nu$ is a kind of anomalous dimension ($\nu=0$ for the free massless case) and $m_{\rm gt}^\prime$ their gluon mass scale. The required value of $m_{\rm gt}^\prime$ \cite[Eq.~(5.31)]{Ngwenya:2025mpo}, at least for the $SU(3)$ case, is an order of magnitude smaller than the $m_{\rm gt}$ estimate of \cite[Eq.~(8)]{Chernodub:2023dok}, obviously related to the quite different fitting function. Moreover, expanding \eqref{CasimirEnergyCF} for large $L$ leads to a behaviour $\sim L^{-3/2}e^{-m_{\rm gt}L}$, also totally different from $\nu\approx0.002$, see \cite[Eq.~(5.30)]{Ngwenya:2025mpo}. We agree with~\cite{Ngwenya:2025cuw} that these observations ask for a deeper understanding of the invoked Casimir fitting functions from an analytical viewpoint. Apart from the global prefactor, the CF model does a good job in supporting the fit~\eqref{CasimirDataFit} of Ref.~\cite{Chernodub:2023dok}.

Now we move on  and start to explicitly compare our findings with the above-mentioned lattice data. Interestingly enough, fitting the $SU(3)$ lattice data~\cite{Duarte:2016iko} for the Landau gauge gluon propagator in four dimensions with the tree level CF propagator gives one a mass $m = 0.54 \, {\rm GeV}$, fairly close to the $m_{\rm gt} = 0.49 \, {\rm GeV}$  value needed to fit the Casimir data of Ref.~\cite{Chernodub:2023dok}. 
To be more precise, we fitted $\frac{Z}{p^2+m^2}$ to the $SU(3)$ lattice data~\cite[$\beta=6.0$, $V=80^4$]{Duarte:2016iko}, where the global $Z$ takes into account the unspecified lattice normalization. In the current approximation, we can safely omit this $Z$ as it merely gives an irrelevant divergent constant contribution, actually vanishing in dimensional regularization, to the vacuum energy. One can also carry out the same analysis for the four-dimensional $SU(2)$ lattice data~\cite[$\beta=2.2$, $V=128^4$]{Cucchieri:2011ig}, finding $m=0.68 \, {\rm GeV}$ in this case. The figures with the generated data points and the tree level CF propagator with best-fit values can be found in App.~\ref{AppCFfits}.

Considering the best-fit mass value $m = 0.54 \, {\rm GeV}$ obtained from the four-dimensional $SU(3)$ gluon propagator~\cite{Duarte:2016iko} as an input, and using the analytical result coming from the CF model for the non-Abelian Casimir energy~\eqref{CasimirEnergyCF}, we can fit the lattice data of Ref.~\cite{Chernodub:2023dok} adjusting only a global multiplicative factor, given by $C = 5.63$. Furthermore, we can also fit the lattice data of Ref.~\cite{Ngwenya:2025cuw} using the same value for the mass as input and adopting a global factor $C = 0.54$. The comparison between our analytical result and the lattice data of Refs.~\cite{Chernodub:2023dok, Ngwenya:2025cuw} is shown in Fig.~\ref{CasimirCF4dSU3Joint}, adopting units of the fundamental string tension $\left(\sqrt{\sigma} = 0.485 \, {\rm GeV}\right)$ to set the scale.

In the same spirit, one can also consider the mass value $m= 0.68 \, {\rm GeV}$ obtained from the four-dimensional $SU(2)$ Landau gauge gluon propagator~\cite{Cucchieri:2011ig} as an input, adjusting only a global multiplicative factor to fit the data of Ref.~\cite{Ngwenya:2025cuw}, given by $C= 0.69$, as is shown in Fig.~\ref{CasimirCF4dSU2}. 

Although this comparison is not intended to be accurate, it is interesting to see that the best-fit mass values for the Casimir energy and the gluon propagator are in the same ballpark at tree-level accuracy. This might suggest that both fitted CF mass scales (Casimir energy vs.~propagators) might actually be the same scale, or at least closely related. This might also indicate that the phenomenological parameter $C_0$, from the CF viewpoint, would correspond to the missing higher order contributions, although its value is approximately 5 times larger than its tree level value for the data~\cite{Chernodub:2023dok}, which does not fit entirely well with a perturbatively stable analysis. This remains a question for future research. Interestingly, the corresponding data of~\cite{Ngwenya:2025cuw}, for the same setups, suggest global scales equaling either $0.54$ or $0.69$, which are far better interpretable from a perturbative viewpoint.\\

Additionally, it would be most interesting if a lattice computation, similar to the one done in Ref.~\cite{Chernodub:2023dok}, could be performed for PMC boundary conditions, to confirm or debunk whether, non-perturbatively, the PMC and PEC boundary conditions yield different Casimir energies.  Note that in the usual (massless) Abelian case, this cannot be the case due to the underlying electric-magnetic duality~\cite{Dudal:PEMC}, but in the non-Abelian case, this duality cannot be maintained once the interactions are switched on, see Ref.~\cite{Deser:1976iy}, leaving also this question to be answered.

\subsection{Parallel wires}

\begin{figure}[t!]
	\begin{minipage}[b]{1.0\linewidth}
\includegraphics[width=\textwidth]{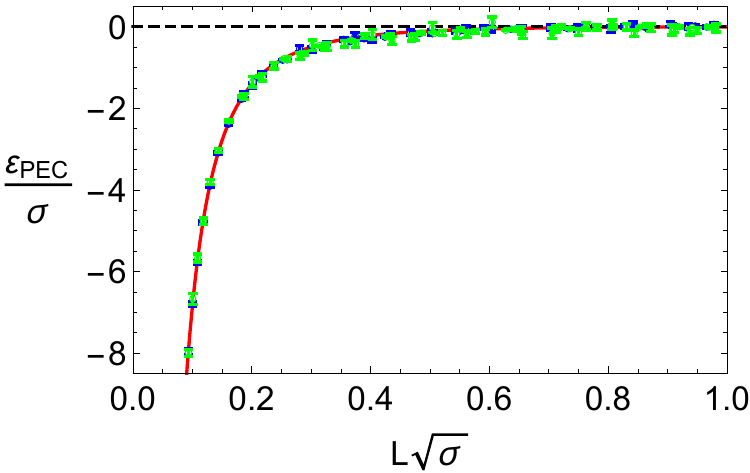}
	\end{minipage} \hfill
\caption{Non-Abelian Casimir energy between parallel wires under PEC boundary conditions for $SU(2)$ gauge group in three dimensions in units of $\sqrt{\sigma}$. The red curve represents the analytical expression~\eqref{CasimirWirePEC}, using $m=0.97 \, {\rm GeV}$ as input and adopting a global factor $C = 1.07$ to fit the lattice data of Ref.~\cite{Chernodub:2018pmt} (blue points) and Ref.~\cite{Ngwenya:2025mpo} (green points).}
\label{CasimirCF3dSU2Joint}
\end{figure}
\begin{figure}[t!]
	\begin{minipage}[b]{1.0\linewidth}
\includegraphics[width=\textwidth]{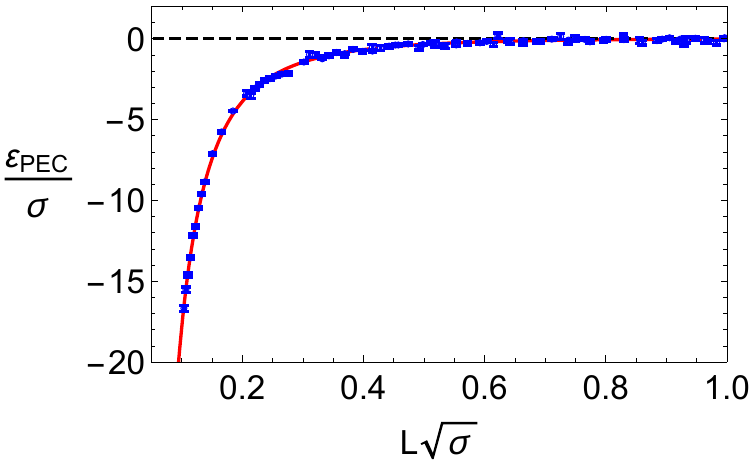}
	\end{minipage} \hfill
\caption{Non-Abelian Casimir energy between parallel wires under PEC boundary conditions for $SU(3)$ gauge group in three dimensions in units of $\sqrt{\sigma}$. The red curve represents the analytical expression~\eqref{CasimirWirePEC}, adopting the mass parameter $m=0.76 \, {\rm GeV}$ and a global factor $C = 1.01$ that were chosen to fit the lattice data of Ref.~\cite{Ngwenya:2025mpo} (in blue).}
\label{CasimirCF3dSU3}
\end{figure}

The first study of the Casimir effect in non-Abelian gauge theory using first principles numerical simulations was done in Ref.~\cite{Chernodub:2018pmt}. The authors investigated the Casimir energy between perfect chromoelectric parallel wires in zero-temperature $SU(2)$ lattice gauge theory in $(2+1)$ dimensions. As in $(3+1)D$, they also reported the emergence of a new massive scale (Casimir mass) that is surprisingly lighter than the lowest glueball mass and characterizes the infrared suppression of the Casimir interaction.

The analytical result for the Casimir energy between  parallel wires under PEC boundary conditions in the CF model is given by Eq.~\eqref{CasimirWirePEC}. As we did before, we can consider the best-fit mass value $m=0.93 \, {\rm GeV}$ from the three-dimensional $SU(2)$ Landau gauge gluon propagator lattice data, generated from Ref.~\cite{Cucchieri:2011ig} (that can be found in App.~\ref{AppCFfits}) as an input, and fit the lattice data for the Casimir energy present in Refs.~\cite{Chernodub:2018pmt,Ngwenya:2025mpo} by adjusting only a global multiplicative factor, given by $C = 1.07$. 
The comparison between our analytical result and the lattice data of Refs.~\cite{Chernodub:2018pmt,Ngwenya:2025mpo} can be found in Fig.~\ref{CasimirCF3dSU2Joint}. Note that for the $SU(2)$ parallel wires, we can nicely fit both sets of lattice data with the same parameters.

One could also fit the data of Refs.~\cite{Chernodub:2018pmt,Ngwenya:2025mpo} using the mass parameter together with a global factor as fitting parameters, without using information from the gluon propagator.  In this case, one would obtain $\left(C=1.05, m=0.86 \, {\rm GeV}\right)$ and $\left(C=1.03, m=0.78 \, {\rm GeV}\right)$ for the data of Refs.~\cite{Chernodub:2018pmt} and~\cite{Ngwenya:2025mpo}, respectively. Note that these are close to the values obtained using the mass coming from the gluon propagator as input, providing very similar plots that will be not shown here.

In a similar fashion, one can also use the mass parameter and a global factor to fit the lattice data of Ref.~\cite{Ngwenya:2025mpo} for parallel wires with $SU(3)$ gauge group. In this case, we find $C=1.01$ and $m=0.76 \, {\rm GeV}$. The comparison between our analytical results and the lattice data~\cite{Ngwenya:2025mpo} can be found in Fig.~\ref{CasimirCF3dSU3}. To our knowledge, there is no lattice data available for the three-dimensional $SU(3)$ Landau gauge gluon propagator, thus we cannot use the mass coming from the gluon propagator as input to repeat the analysis done for the other cases. 

It is important to stress that in the parallel wire case we remarkably find $C \approx 1$. Since this global factor was taking into account all missing contributions, this suggests that in the parallel wire case, the CF model seems to capture the essential physics already at tree level accuracy.

\section{Concluding Remarks}\label{SecConclusions}

We investigated the non-Abelian Casimir energy from a boundary QFT perspective, employing the CF Lagrangian as a simple effective model to describe the infrared regime of YM theory. The configuration of interest consists of two parallel plates (or wires), separated by a distance $L$ from each other, subjected to either PMC or PEC boundary conditions. While our analysis focuses on this specific geometry, the techniques developed here are readily extendable to more general setups. 

Introducing auxiliary boundary fields, we obtained an effective boundary action, at leading order, to analyze the dynamics of the system, using it to compute the Casimir energy through two different methods: first, directly from the path integral, and second, via the EMT. Both methods yielded identical results. Comparing the results for PMC and PEC boundary conditions, we found that they differ by a global multiplicative factor, in contrast with the massless Abelian case, where they are equal. This difference in behavior gives rise to a van Dam--Veltman--Zakharov discontinuity in the massless limit.

Noteworthy, our findings are compatible with recent lattice data, providing the correct analytical structure to match the data, up to a global multiplicative factor. Remarkably, for the parallel wires case, this factor is close to 1, suggesting that in this case the CF model seems to capture the essential physics already at tree-level accuracy. Interestingly enough, the best-fit values for the Casimir energy and the gluon propagator are in the same ballpark, indicating that these mass scales might be the same, or at least closely related. It would be highly interesting to extend the analysis beyond leading order to probe the loop effects of both massive gluons and massless ghosts on the Casimir energy. Likewise, it would be nice if we could explore the ``glueton'' boundary states of \cite{Chernodub:2023dok} also from an analytical perspective. This being said, the effective action for our boundary fields, see Eqs.~\eqref{SbcompactwithK}-\eqref{KinvEandF}, does imply a Euclidean mass singularity at $\mathbf k^2=-m^2$, that is, where $\omega_k=\sqrt{\mathbf k^2+m^2}=0$ in the boundary fields propagator matrix, which might suggest that the boundary fields are perhaps closely related to these ``gluetons'', with their mass scale dictated by $m$.

Finally, we note that it is well-known that the non-perturbative regime of YM theory is plagued by Gribov copies~\cite{Gribov78}, jeopardizing the FP method and demanding its improvement in order to handle them in the infrared. To deal with this problem, one could restrict the functional integral to a region which is crossed by all gauge orbits, but free of (infinitesimal) Gribov copies. The boundary of this region in field space is the so-called Gribov horizon. The implementation of this idea was first done at leading order~\cite{Gribov78}, then extended to all orders~\cite{Zwanziger89}, and later refined to account for other non-perturbative effects~\cite{Dudal:2008sp}. Notably, a non-perturbative mass scale also emerges from the restriction to the Gribov horizon\footnote{It has been argued that the mass of the CF model gives another way to partially deal with the gauge fixing ambiguity, see \cite{Serreau:2012cg,Pelaez21} and references therein.}. 
Extending the techniques from this paper to the Gribov-Zwanziger framework is work in progress, we will come back to this in a forthcoming paper.

\section*{Note added}
Very shortly after finalizing this paper, \cite{Karabali:2025olx} appeared, reporting similar findings.

\section*{Acknowledgments}

The authors wish to thank K.~Boon for useful discussions in the lead-up to this paper. The authors are grateful to M.~Chernodub, B.~A.~Ngwenya, O.~Oliveira, and P.~Silva for providing the lattice data used in this work, and to the authors of~\cite{Karabali:2025olx} for useful discussions related to their and our work.
The work of D.~Dudal was supported by KU Leuven IF project C14/21/087. The work of S.~Stouten was funded by FWO PhD-fellowship fundamental research (file number: 1132823N). P.~De~Fabritiis thanks the National Council for Scientific and Technological Development – CNPq for the financial support (No. 402459/2024-5).

\appendix 

\section{Tree-level fits for the gluon propagator} \label{AppCFfits}

\begin{figure}[t!]
	\begin{minipage}[b]{1.0\linewidth}
\includegraphics[width=\textwidth]{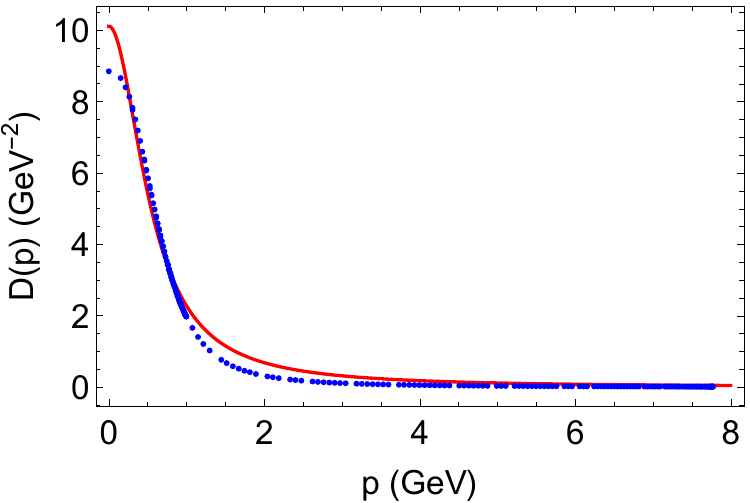}
	\end{minipage} \hfill
	\caption{Lattice data~\cite{Duarte:2016iko} for the four-dimensional $SU(3)$ gluon propagator in Landau gauge with $\beta = 6.0$ and $V = 80^4$, together with fitted tree-level CF form factor~\eqref{CFTreeLevelFit}, with values $Z = 2.93$ and $m = 0.54 \, {\rm GeV}$.}
\label{GluonPropSU3CFtreeNew}
\end{figure}

    \begin{figure}[t!]
	\begin{minipage}[b]{1.0\linewidth}
\includegraphics[width=\textwidth]{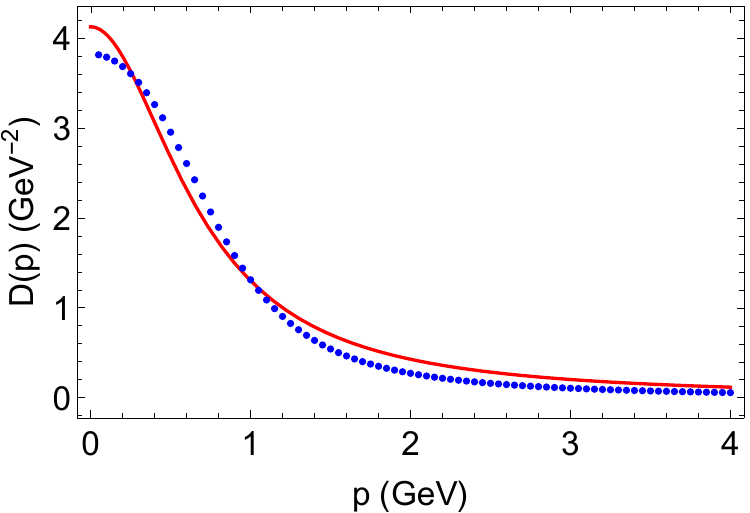}
	\end{minipage} \hfill
	\caption{ Data points for the four-dimensional $SU(2)$ gluon propagator in the Landau gauge with $\beta = 2.2$ and $V = 128^4$, generated from Ref.~\cite{Cucchieri:2011ig} (see main text for more details), together with fitted tree-level CF form factor~\eqref{CFTreeLevelFit}, with values $Z = 1.90$ and $m = 0.68 \, {\rm GeV}$.}
\label{GluonProp4DSU2CFtreeNew2}
\end{figure}


In this Appendix, we provide the lattice data and the best-fit values obtained from the \verb"Mathematica" function ``FindFit" for the gluon propagator using the tree-level CF propagator (with a global factor) as the fitting profile:
\begin{align} \label{CFTreeLevelFit}
    D(p) = \frac{Z}{p^2 + m^2}.
\end{align}

The lattice data for the four-dimensional $SU(3)$ gluon propagator in the Landau gauge can be found in Ref.~\cite{Duarte:2016iko}, and is shown in Fig.~\ref{GluonPropSU3CFtreeNew} with the plot of Eq.~\eqref{CFTreeLevelFit} with best-fit values $\left(Z= 1.44, m=0.67  \, {\rm GeV}\right)$.  Error bars are not included here, being virtually invisible on the plot.

For the $SU(2)$ case we used Ref.~\cite{Cucchieri:2011ig} to generate a set of data points used here for both four-dimensional and three-dimensional gluon propagator in the Landau gauge. These points are generated with 
the fitting function of the form (Eq.~28 of Ref.~\cite{Cucchieri:2011ig}):
\begin{align} \label{FitFunctionF4}
f_4(p^2) = C \frac{p^4 + (s+1) p^2 + s}{p^6 + (k + u^2) p^4 + (k u^2 + t^2) p^2 + k t^2}.   
\end{align}
In the four-dimensional case, the coefficients can be obtained from Table~VI ($V = 128^4$, $\beta = 2.2$), being given by: $ \{ C=0.793, u= 0.727, t= 0.791, s= 1.903, k= 0.631 \}$. We generated $80$ equidistant points between $0.05$ and $4$ GeV using Eq.~\eqref{FitFunctionF4} with the coefficients above. These points are shown in Fig.~\ref{GluonProp4DSU2CFtreeNew2}, together with the plot of Eq.~\eqref{CFTreeLevelFit} with best-fit values $Z= 1.90$ and $m=0.68 \, {\rm GeV}$.

\begin{figure}[t!]
	\begin{minipage}[b]{1.0\linewidth}
\includegraphics[width=\textwidth]{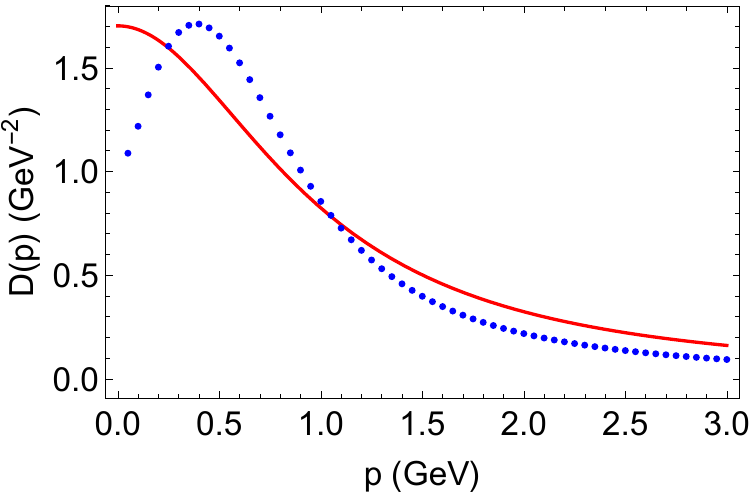}
	\end{minipage} \hfill
	\caption{Data points for the three-dimensional $SU(2)$ gluon propagator in the Landau gauge with $\beta = 3.0$ and $V = 320^3$, generated from Ref.~\cite{Cucchieri:2011ig} (see main text for more details), together with fitted tree-level CF form factor~\eqref{CFTreeLevelFit}, with values $Z = 1.60$ and $m = 0.97 \, {\rm GeV}$.}
\label{GluonProp3DSU2CFtree}
\end{figure}

For the three-dimensional case, the coefficients can be extracted from Table~IX ($V = 320^3$, $\beta = 3.0$), being given by: $\{C=0.408, u=0.656, t=0.619, s=0.023, k=0.046\}$. We generated $60$ equidistant points between $0.05$ and $3$ GeV using Eq.~\eqref{FitFunctionF4} with the coefficients above. These points are shown in Fig.~\ref{GluonProp3DSU2CFtree}, together with the plot of Eq.~\eqref{CFTreeLevelFit} with best-fit values $Z= 1.60$ and $m=0.97 \, {\rm GeV}$.


\bibliography{bibliography}

\end{document}